\documentclass[12pt]{article} 

\usepackage{amsmath}
\usepackage{amssymb}
\usepackage{amsfonts}
\usepackage{amsthm}
\usepackage{mathtools}
\usepackage{braket}
\usepackage{dirtytalk}
\usepackage{dsfont}
\usepackage{cancel}
\usepackage{bbm}
\usepackage{subcaption}
\usepackage{tikz-cd}
\usepackage{booktabs}

\newtheorem{theorem}{Theorem}
\newtheorem{proposition}{Proposition}
\newtheorem{definition}{Definition}

\usepackage[style=numeric-comp, sorting=none, bibstyle=ieee]{biblatex}
\usepackage[colorlinks, allcolors=blue]{hyperref}
\bibliography{biblio}

\newcommand{\GL}{\mathrm{GL}}
\newcommand{\Or}{\mathrm{O}}

\newcommand{\SO}{\mathrm{SO}}
\newcommand{\SU}{\mathrm{SU}}
\newcommand{\U}{\mathrm{U}}

\newcommand{\Z}{\mathbb{Z}}

\newcommand{\R}{\mathbb{R}}
\newcommand{\C}{\mathbb{C}}

\newcommand{\pqty}[1]{\left( #1 \right)}
\newcommand{\bqty}[1]{\left[ #1 \right]}
\newcommand{\abs}[1]{\left\lvert #1\right\rvert}
\newcommand{\expval}[1]{\langle #1\rangle}

\newcommand {\dv}[3][ ]{
	\ifx #1 { }
	\frac{d #2}{d #3}
	\else
	\frac{d^{#1} #2}{d #3^{#1}}
	\fi
}
\newcommand {\pdv}[3][ ]{
	\ifx #1 { }
	\frac{\partial #2}{\partial #3}
	\else
	\frac{\partial^{#1} #2}{\partial #3^{#1}}
	\fi
}
\newcommand {\fdv}[3][ ]{
	\ifx #1 { }
	\frac{\delta #2}{\delta #3}
	\else
	\frac{\delta^{#1} #2}{\delta #3^{#1}}
	\fi
}

\newcommand{\tr}{\operatorname{tr}}

\renewcommand{\Re}{\operatorname{Re}}

\newcommand{\Htot}{\mathcal{H}_{\mathrm{tot}}}
\newcommand{\Hphys}{\mathcal{H}_{\mathrm{phys}}}

\title{Counting gauge-invariant states with \\ matter fields and finite gauge groups}

\author{A. Mariani,\\
	Physics Department, University of Turin \& INFN, Turin unit,\\ Via Pietro Giuria 1, I-10125 Turin, Italy \\ \\}

\date{\today}

\begin{document}
	
	\maketitle
	
	\begin{abstract}
		Gauge theories with finite gauge groups have applications to quantum simulation and quantum gravity. Recently, the exact number of gauge-invariant states was computed for pure gauge theories on arbitrary lattices. In this work, we generalize this counting to include the case of scalar and fermionic matter, as well as various kinds of boundary conditions. As a byproduct, we consider several related questions, such as the implementation of charge conjugation for a generic finite group. These results are relevant for resource estimation and also as a crosscheck when working in a gauge-invariant basis.
	\end{abstract}
	
	\tableofcontents

	\section{Introduction}
	
	Advances in quantum computing have renewed interest in the Hamiltonian formulation of lattice gauge theories. One of its key theoretical challenges is that bosonic quantum field theories generically have a locally infinite-dimensional Hilbert space, which must be mapped to a finite-dimensional quantum register. Many solutions to this problem have been proposed. Here we focus on one such solution, which involves replacing the gauge group (a compact Lie group) with one of its finite subgroups \cite{Hasenfratz1, ZoharBurrello, Ercoetal1, Ercoetal2, Alexandru:2019nsa}.
	
	Theories with a finite gauge group should be understood as effective theories with a finite cut-off, which can however be increased via improvement \cite{Hasenfratz1}. This has been demonstrated for the largest crystal-like subgroup of $\SU(3)$ \cite{Alexandru:2019nsa, S1080}. Basic quantum routines have also been constructed for several finite subgroups of $\SU(2)$ and $\SU(3)$ \cite{BinaryTetrahedral, BinaryOctahedral,Gustafson:2024kym,Murairi:2024xpc}.
	
	We also note that gauge theories with a finite gauge group also have intrinsic interest in condensed matter \cite{GiantFluctuations, DualGaugeTheory2D,DualGaugeTheory3D,CondMatDiscreteGaugeTheory2,Manjunath_2021} and quantum gravity \cite{HarlowOoguri}, including with matter fields \cite{Toolkit}.
	
	A peculiarity of gauge theories is that only a small subset of the Hilbert space represents physical states. The physical states are those which are gauge-invariant and satisfy the so-called Gauss law constraint. There is a trade-off in working directly in the gauge-invariant subspace: while fewer qubits are required, operations are more complex. It is not yet clear whether it is worth the trade-off, and the answer may be hardware-dependent. Many proposals work directly with gauge-invariant variables, for example recently \cite{Maiti:2024jwk, Chandrasekharan:2025smw, Grabowska:2024emw, Burbano:2024uvn, Raychowdhury:2018osk, Raychowdhury:2019iki, Kadam:2022ipf, Wang:2024dda} (not an exhaustive list), although others argue against this \cite{Carena:2024dzu}. For finite groups, gauge-invariant formulations have been considered in terms of spin networks \cite{MPE} as well as holonomies \cite{Mariani:2024osg} (see also \cite{Grabowska:2024emw,Burbano:2024uvn} for closely related ideas on this last point). 
	
	An important question when working in the gauge-invariant subspace is the count of the dimension of the physical subspace. Interest in this question relates first of all to resource estimation, i.e.\ one wants to know the expected memory savings when working with gauge-invariant variables. Moreover, the exact count of the dimension is also important for cross-checking that one has indeed produced all the gauge-invariant states when working with an explicit representation. Counts of gauge-invariant states have played a role for example in \cite{Ballini:2024wbe,Biswas:2022env,Sau:2023clm,Sau:2024uur} (not an exhaustive list).
	
	The problem of counting the dimension of the gauge-invariant subspace for finite gauge groups was recently solved in \cite{MPE, Mariani:2024osg} in the case of pure gauge theories on arbitrary lattices. In this work, we extend those results in several directions, mainly by including the case of scalar and fermionic matter. Moreover, we include a discussion of twisted boundary conditions which was absent from previous work. These results we discuss in Section \ref{sec:general}, including general counting formulas. As part of our discussion of twisted boundary conditions, in Section \ref{sec:charge conjugation} we consider the problem of defining charge conjugation for finite gauge groups. We hope to provide the reader with a sufficiently powerful toolkit to approach this problem in whatever specific situation might be useful for them. Moreover, several of our results are of interest also for compact Lie groups. Since matter fields have peculiarities of their own, we then discuss scalars and fermions separately. In Section \ref{sec:scalar} we discuss scalar fields with a finite-dimensional Hilbert space and arbitrary group representation. Then in Section \ref{sec:fermions} we discuss fermionic matter and how our results change when accounting for the method used to overcome fermion doubling.
	
	The reader who is solely interested in the final results can jump directly to the relevant formulas. These are in particular eq.\eqref{eq:master counting formula} for the general counting formula, instantiated in the cases of scalars in eq.\eqref{eq:counting formula scalars} and of fermions in eq.\eqref{eq:fermion gauge-invariant states number}. For some twisted boundary conditions, the relevant formula is instead given in eqs.\eqref{eq:master formula twisted non homomorphism}-\eqref{eq:master formula twisted}. In the conclusions, we have also included a summary of the various formulas we have obtained. 
	
	\section{General construction}\label{sec:general}
	
	In this section, we first discuss the pure gauge sector, the Hamiltonian and the action of gauge transformations. We then give a general formula for the number of gauge-invariant states with arbitrary matter fields. This formula is rather general and will be instantiated in the cases of scalar and fermionic matter fields in Sections \ref{sec:scalar} and \ref{sec:fermions} respectively. We then discuss how to modify this formula in the case of twisted boundary conditions, and give several examples of applications. As it is relevant to this discussion, we conclude by discussing how charge conjugation is implemented for general gauge groups.
	
	\subsection{The pure gauge sector}
	
	In the Hamiltonian formulation of lattice field theory, time is continuous while space is discretized as a finite set of points. A lattice $X$ consists of such a set of points called \say{sites}, together with \say{links}, i.e. intervals connecting sites, as well as faces bounded by links, etc. In mathematical language, it is a \say{cell complex}. The standard example is a periodic hypercubic lattice, but most of our results are valid for arbitrary lattices. For questions of gauge invariance only sites and links are required. Therefore the lattice $X$ can be seen as an arbitrary graph with vertices being the sites and edges being the links. The graph is allowed to have multiple edges and self-loops, as long as each edge is counted separately\footnote{A spatial graph with multiple edges was in fact considered in \cite{MPE}, where the formula for the number of gauge-invariant states was explicitly checked to hold.  On the other hand, in \cite{Mariani:2024osg} gauge-fixing on a maximal tree was shown to reduce the spatial graph to one with one vertex and $E-V+1$ self-loops, and the same formula was also shown to hold in that case.}. This is the type of lattice we will work with throughout this paper. We will denote with $V$ the number of sites and with $E$ the number of links (note this is different from \cite{MPE,Mariani:2024osg}, where the number of links was denoted as $L$).
	
	The total Hilbert space of the theory is given by the tensor product of the gauge degrees of freedom with the matter degrees of freedom,
	\begin{equation}
		\Htot = \mathcal{H}_{G} \otimes \mathcal{H}_{M} \ .
	\end{equation}
	The matter Hilbert space can be factorized as a tensor product of local Hilbert spaces associated to sites,
	\begin{equation}
		\mathcal{H}_{M} = \bigotimes_{x \in \mathrm{sites}} (\mathcal{H}_{M})_x \ .
	\end{equation}
	The details of this decomposition depend on the type of matter and are further elaborated in Sections \ref{sec:scalar} and \ref{sec:fermions} respectively for scalars and fermions.
	
	The gauge field instead lives on lattice links, and therefore its Hilbert space can be split as a tensor product over links,
	\begin{equation}
		\mathcal{H}_{G} =\bigotimes_{l \in \mathrm{links}} (\mathcal{H}_{G})_l \ .
	\end{equation}
	In fact, a classical gauge field configuration is given by an assignment of a group element $g_l$ to each lattice link $l$. Here we assume that lattice links carry an arbitrary orientation; when a link is traversed opposite its orientation, $g_l$ is replaced by $g_l^{-1}$. Thus, classically, a gauge field configuration is an element of $G^E$, i.e. one copy of the gauge group per lattice link ($E$ being the number of lattice links). The quantum Hilbert space can be thought of as being the space of square-integrable wavefunctions $\psi: G^E \to \C$ which assign to each classical configuration its quantum probability amplitude. Thus the quantum Hilbert space is
	\begin{equation}
		\mathcal{H}_{G} =L^2(G^E) \cong \bigotimes_{l \in \mathrm{links}} L^2(G) \ ,
	\end{equation}
	where $L^2(G)$ is the space of square-integrable functions on the gauge group $G$. As explained in the introduction, we are interested in the situation where the gauge group is a finite group. In this case, functions on $G$ are automatically square-integrable and the local gauge Hilbert space is isomorphic to the complex vector space $\C[G] \cong \C^{\abs{G}}$ of dimension $\abs{G}$. Thus 
	\begin{equation}
		\mathcal{H}_{G} = \bigotimes_{l \in \mathrm{links}} \C[G] \ .
	\end{equation}
	With this notation, we emphasize that $\C[G]$ has a privileged basis $\{\ket{g} \}$ indexed by group elements $g \in G$. The states in this basis are orthonormal, i.e. $\expval{g | h} = \delta_{gh}$. By extension, an orthonormal basis for $\mathcal{H}_{G}$ is given by
	\begin{equation}
		\ket{\{g_l\}} \equiv \bigotimes_{l \in \mathrm{links}} \ket{g_l} \ .
	\end{equation}
	Thus we see that
	\begin{equation}
		\dim \mathcal{H}_{G} = \abs{G}^E \ .
	\end{equation}
	Two important operators on the local gauge Hilbert space are left and right translations $L_g$ and $R_g$, which act as
	\begin{equation}
		\label{eq:left and right translations}
		L_g \ket{g'} \equiv \ket{g g'} \ , \qquad R_g \ket{g'} \equiv \ket{g' g^{-1}} \ . 
	\end{equation}
	It is not hard to show that $L_g$ and $R_g$ are unitary representations of the gauge group $G$ (see also \cite{Serre, KnappLieGroups}).
	
	A key feature of gauge theories is that their Hilbert space must be invariant under all gauge transformations. This is known as the \say{Gauss law} constraint. A gauge transformation is given by an assignment $\{g_x\}$ of a group element $g_x \in G$ to each lattice site $x$. Such a gauge transformation acts on the Hilbert space via a unitary operator $\mathcal{G}(\{g_x\})$ defined as
	\begin{equation}
		\label{eq:gauge transformation generic}
		\mathcal{G}(\{g_x\}) = \bigotimes_{l = \expval{xy} \in \mathrm{links}} L_{g_x} R_{g_y} \bigotimes_{x \in \mathrm{sites}} U_{g_x} \ .
	\end{equation}
	Here $U_g$ is a unitary group representation which implements the gauge transformation on each local matter Hilbert space. Its exact form differs between scalars and fermions and will be discussed in their dedicated section. The action of the gauge transformations on the gauge sector is just to send the group element $g_l$ on the link $l =\expval{xy}$ starting at $x$ and ending at $y$ to the group element $g_l' = g_x g_l g_y^{-1}$. Since the independent degrees of freedom of gauge transformations are site-based, sometimes it is useful to factor $\mathcal{G}(\{g_x\})$ into its independent components as
	\begin{equation}
		\label{eq:gauge transformation site based}
		\mathcal{G}(\{g_x\}) = \bigotimes_{x\in \mathrm{sites}} \mathcal{G}(g_x) \ , \qquad \mathcal{G}(g_x) =\bigotimes_{x=l_-} L_{g_x} \bigotimes_{x=l_+} R_{g_x} \otimes U_{g_x} \ ,
	\end{equation}
	where the first two terms refer to the gauge sector and the third to the matter sector. Here $l_\pm$ are the end and start vertex of link $l$, so we left multiply by $g_x$ on all links starting at $x$ and right multiply by $g_x^{-1}$ on all links ending at $x$. Thus eq.\eqref{eq:gauge transformation site based} is just a different way of writing down the action of eq.\eqref{eq:gauge transformation generic}. Whether $\mathcal{G}$ refers to the full gauge transformation or to the one based on the site is clear from context and from its argument. 
	
	Note that the gauge transformations form a group, i.e. $\mathcal{G}(\{g_x\}) \mathcal{G}(\{g_x'\})=\mathcal{G}(\{g_x g_x'\})$. Assuming that $U$ is faithful, then the group of gauge transformations is isomorphic to $G^V$ (i.e. $V$ copies of the gauge group) \footnote{Note that this is unlike in the pure gauge case, where one had to quotient out the \say{diagonal} center subgroup \cite{Mariani:2024osg}.}.
	
	The physical Hilbert space $\Hphys$ of the gauge theory is the subspace of $\Htot$ which is invariant under all gauge transformations \cite{KogSuss, Osborne, Tong}, i.e. it is made of those states which satisfy  
	\begin{equation}
		\label{eq:physical state definition}
		\mathcal{G}(\{g_x\}) \ket{\psi} = \ket{\psi} \ ,
	\end{equation}
	for any possible assignment $\{g_x\}$ of group elements $g_x$ to sites $x$. Sectors where the Gauss law eq.\eqref{eq:physical state definition} is violated at one or more sites are also physical and represent the presence of static charges. We will consider this possibility in Section \ref{sec:nontrivial sectors}.
	
	The Hamiltonian of the theory is given by the sum of a gauge part and a matter part, 
	\begin{equation}
		H = H_G + H_M \ .
	\end{equation}
	The matter Hamiltonian includes the coupling to the gauge fields, and depends on the exact models. We will discuss several possible models in Sections \ref{sec:scalar} and \ref{sec:fermions}. The gauge field Hamiltonian is given by the sum of an electric and a magnetic part
	\cite{KogSuss, Orland, ZoharBurrello, Caspar_Wiese, HarlowOoguri, MPE},
	\begin{equation}\label{eq:gauge hamiltonian}
		H_G = \lambda_E \sum_{l \in \mathrm{links}} h_E + \lambda_B \sum_{\square} h_B(g_\square) \ .
	\end{equation}
	In this formula $g_\square$ is a plaquette variable and $h_B$ is a class function, i.e.\ it satisfies $h_B(hgh^{-1}) = h_B(g)$. On a more general (non-hypercubic) lattice, $\square$ would be the $2$-cells. The electric term acts on each link as a Laplace operator on the finite group,
	\begin{equation}
		\label{eq:group Laplacian}
		h_E = \sum_{k \in \Gamma} (1-L_k) \ ,
	\end{equation}
	where $L_k$ was defined in eq.\eqref{eq:left and right translations} and $\Gamma \subset G$ is a subset of the gauge group invariant under inversion and conjugation \cite{MPE}. For our purposes the only relevant property of the gauge Hamiltonian is that it is invariant under the gauge transformations \eqref{eq:gauge transformation generic}. For further discussion of its properties, we refer the reader to \cite{MPE}. 
	
	\subsection{Master counting formula}\label{sec:master counting}
	
	In this section, we derive a formula for the number of gauge-invariant states under fairly generic assumptions. This extends the results of \cite{MPE, Mariani:2024osg} which were limited to the pure gauge case. Moreover, the derivation in this section improves on previous techniques in two ways: first of all, we do not make use of representation theory, which makes the derivation more accessible. Secondly, it does not require an explicit description of the physical Hilbert space, which allows the calculation in cases where such a description is more complicated than in pure gauge, for example with matter fields or twisted boundaries. The formula derived in this section is valid for arbitrary lattice geometry, unless twisted boundary conditions are imposed. We deal with that case in Section \ref{sec:twisted}. 
	
	As we have seen in the previous section, the total Hilbert space is given by the tensor product of local gauge Hilbert spaces on lattice links, and local matter Hilbert spaces on lattice sites,
	\begin{equation}
		\label{eq:total Hilbert space}
		\Htot = \bigotimes_{l \in \mathrm{links} } \C[G]_l  \bigotimes_{x \in \mathrm{sites}} (\mathcal{H}_M)_x \ . 
	\end{equation}
	The action of gauge transformations is given in eq.\eqref{eq:gauge transformation generic} or eq.\eqref{eq:gauge transformation site based}. The general strategy to compute the number of gauge-invariant states proceeds as follows. One first constructs the projector $P: \Htot \to \Hphys$ onto the gauge-invariant sector,
	\begin{equation}
		\label{eq:projector onto Hphys}
		P = \frac{1}{\abs{G}^V} \sum_{\{g_x\}} \mathcal{G}(\{g_x\}) \ .
	\end{equation}
	The projector is just the average over all gauge transformations. Since the gauge transformations form a group, i.e. $\mathcal{G}(\{g_x\}) \mathcal{G}(\{g_x'\}) = \mathcal{G}(\{g_x g'_x\})$, the projector is correctly normalized so that $P^2=P$. This implies that it has eigenvalues equal to either $1$ (corresponding to gauge-invariant states) or $0$ (states which are \textit{not} gauge-invariant). Then one obtains the number of gauge-invariant states by summing up the $1$ eigenvalues, i.e. by taking the trace:
	\begin{equation}
		\label{eq:summing up}
		\dim \Hphys = \tr P = \frac{1}{\abs{G}^V} \sum_{\{g_x\}} \tr \mathcal{G}(\{g_x\}) \ .
	\end{equation}
	To proceed, we now compute the trace of an arbitrary gauge transformation. From the general form of gauge transformations, eq.\eqref{eq:gauge transformation generic}, and the fact that the trace of a tensor product is the product of the traces, one has
	\begin{equation}
		\tr\mathcal{G}(\{g_x\}) = \prod_{l = \expval{xy} \in \mathrm{links}} \tr (L_{g_x} R_{g_y}) \prod_x \chi_U(g_x) \ ,
	\end{equation}
	where $\chi_U=\tr U$ is the character of $U$. No further simplification is possible in the matter term. The trace of the gauge term instead can be computed by inserting an orthonormal basis of states $\{\ket{g}\}$ of the local link Hilbert space,
	\begin{equation}
		\tr (L_{g_x} R_{g_y}) = \sum_{g\in G} \bra{g} L_{g_x} R_{g_y} \ket{g} = \sum_{g \in G} \expval{g | g_x g g_y^{-1}} =\sum_{g \in G} \expval{g_y | g^{-1} g_x g} \ .
	\end{equation}
	where in the last term we could shuffle around some group elements owing to the orthonormality of the basis, i.e. $\expval{g|h}=\delta_{gh}$. From this last expression, it is clear that the trace is going to be zero unless $g_x$ and $g_y$ are conjugate. On the other hand, if this is the case, then we can write $g_x = h^{-1} g_y h$ for some $h \in G$ and substitute to get 
	\begin{equation}
		\sum_{g}\braket{g_y | g^{-1} g_x g }  = \sum_{g}\braket{g_y | (hg)^{-1} g_y (hg) } = \sum_{g}\braket{g_y | g^{-1} g_y g } \ ,
	\end{equation}
	since summing over $g$ is equivalent to summing over $hg$. Therefore we see that this sum just counts the number of $g \in G$ which commute with $g_y$. The set of such elements is known as the centralizer $C_G(g_y)$ \cite{Serre}. The orbit-stabilizer theorem implies that \cite{Serre}
	\begin{equation}
		\abs{C_G(g)} = \frac{\abs{G}}{\abs{C(g)}} \ ,
	\end{equation}
	where $C(g)$ is the conjugacy class to which $g$ belongs. Putting these results together, we see that 
	\begin{equation}
		\tr (L_{g_x} R_{g_y}) = \begin{cases} \frac{\abs{G}}{\abs{C}} & \mathrm{if}\, g_x,g_y \in C \, \mathrm{conjugate} \\ 0 & \mathrm{otherwise}\end{cases}
	\end{equation}
	Therefore, since the lattice is assumed to be connected, to obtain a non-zero result all $g_x$ must be in the same conjugacy class, so that one gets
	\begin{equation}
		\tr\mathcal{G}(\{g_x\}) = \begin{cases} \pqty{\frac{G}{\abs{C}}}^E \chi_U(C)^V & \mathrm{all}\, g_x \in C \, \mathrm{conjugate} \\ 0 & \mathrm{otherwise}\end{cases}
	\end{equation}
	since $\chi_U(g)$, being a character, depends only on the conjugacy class of $g$. Therefore substituting into the trace of the projector we get
	\begin{equation}
		\label{eq:intermediate Hphys calculation}
		\dim \Hphys = \frac{1}{\abs{G}^V} \sum_C \sum_{\{g_x\} \in C} \pqty{\frac{\abs{G}}{\abs{C}}}^E \chi_U(C)^V = \sum_{C} \pqty{\frac{\abs{G}}{\abs{C}}}^{E-V} \chi_U(C)^V \ ,
	\end{equation}
	where the sum is over all conjugacy classes $C$ of $G$.
	
	We now state this result more clearly. Let $G$ be an arbitrary finite gauge group. On an arbitrary lattice with $V$ sites and $E$ edges, with gauge transformations given by eq.\eqref{eq:gauge transformation generic} which act the same on every link and every site, and the matter field representation having character $\chi_U$, the number of gauge-invariant states is given by
	\begin{equation}
		\label{eq:master counting formula}
		\dim \Hphys = \sum_{C} \pqty{\frac{\abs{G}}{\abs{C}}}^{E-V} \chi_U(C)^V \ .
	\end{equation}
	This formula is rather general, but it does not apply in all cases of interest. For example, the assumption that the Gauss law is the same at every link is violated in the case of twisted boundary conditions. We consider this case in more detail in Section \ref{sec:twisted}. Alternatively, it is also possible to gauge only a subgroup $H < G$ of the gauge group; the formula remains essentially unchanged (with $H$ replacing $G$ everywhere in eq.\eqref{eq:master counting formula}) but with some interesting implications for the global symmetry (see Section \ref{sec:scalar examples}). It is also possible and sometimes useful to implement further constraints on the physical Hilbert space apart from the Gauss law constraint. If these further constraints are implemented by a projector $P'$, then the number of gauge-invariant states satisfying these extra constraints can be computed as $\tr(PP')$. We will see an example of such a calculation in Section \ref{sec:superselection}. Finally, it might also be interesting to consider the case where the matter fields transform under a different gauge group representation $U_x$ at each site $x$. By repeating the above calculation, it is easy to see that the corresponding result in this case is
	\begin{equation}
		\label{eq:master formula different sites}
		\dim \Hphys = \sum_{C} \pqty{\frac{\abs{G}}{\abs{C}}}^{E-V} \prod_{x \in \mathrm{sites}} \chi_{U_x}(C) \ .
	\end{equation}
	We will make use of this formula in the next section. 
	
	\subsection{Non-trivial Gauss law sectors}\label{sec:nontrivial sectors}
	
	As discussed in the previous sections, sectors where the Gauss law is violated at one or more points are also physical and they correspond to the insertion of static charges. In this section, we limit ourselves to pure gauge theory.
	
	The total Hilbert space $\Htot$ carries a (non-irreducible) representation of the group of gauge transformations $G^V$ (i.e. $V$ copies of the gauge group $G$, one per site), given by the linear action eq.\eqref{eq:gauge transformation generic}. In general, given a finite or compact Lie group $H$ and an arbitrary representation $\rho$ of $H$, one can decompose $\rho$ in terms of irreps $\rho_q$ of $H$,
	\begin{equation}
		\rho \cong \bigoplus_q (\rho_q)^{n_q} \ ,
	\end{equation}
	where $n_q$ is the multiplicity of $\rho_q$ in $\rho$. Taking traces, one obtains for the characters $\chi=\tr \rho$ and $\chi_q=\tr\rho_q$,
	\begin{equation}
		\chi = \sum_q n_q \chi_q \ .
	\end{equation} 
	The character orthogonality relations then give the following formula for the multiplicities,
	\begin{equation}
		n_q = \frac{1}{\abs{H}} \sum_{h \in H} \chi_q(h)^* \chi(h) \ .
	\end{equation}
	The projector $P_q: \rho \to (\rho_q)^{n_q}$ onto the subspace trasforming as $\rho_q$ (also known as the $q$-isotypic component) is given by (see e.g. Theorem 8 in \cite{Serre})
	\begin{equation}
		P_q = \frac{\dim q}{\abs{H}} \sum_{h \in H} \chi_q(h)^* \rho(h) \ .
	\end{equation}
	Note that this is correctly normalized, i.e. $(P_q)^2 = P_q$. Then the number of states in the subspace $(\rho_q)^{n_q}$ is given by
	\begin{equation}
		\label{eq:isotypic projector}
		\tr(P_q) = \frac{\dim q}{\abs{H}} \sum_{h \in H} \chi_q(h)^* \chi(h) = n_q \dim q \ , 
	\end{equation}
	as expected. In our case, the relevant group is $H=G^V$ and any irrep of $G^V$ is given by a tensor product of irreps of $G$ (one per site). The physical Hilbert space eq.\eqref{eq:physical state definition} corresponds to choosing the trivial irrep at each site. Generically one then has the decomposition of the total Hilbert space
	\begin{equation}
		\Htot = \bigoplus_{\{q\}} \mathcal{H}_{\{q\}} \ ,
	\end{equation}
	where $\{q\}$ is an assignment of an irrep $q_x$ of $G$ to each site $x$ of the lattice, i.e. a distribution of static charges. Here $\mathcal{H}_{\{q\}}$ is the subspace transforming as $\{q\}$. Then the projector $P_{\{q\}}: \Htot \to \mathcal{H}_{\{q\}}$ is given by plugging $H=G^V$ in eq.\eqref{eq:isotypic projector}, i.e.
	\begin{equation}
		P_{\{q\}} = \frac{\prod_x \dim q_x}{\abs{G}^V} \sum_{\{g_x\} \in G^V} \pqty{\prod_x\chi_{q_x}(g_x)^*} \mathcal{G}(\{g_x\}) \ .
	\end{equation}
	In the case where $q_x$ is the trivial representation at each site, this reduces to the projector onto gauge-invariant states eq.\eqref{eq:projector onto Hphys}. Then the dimension of the subspace $\mathcal{H}_{\{q\}}$ is given by the trace of the projector; the resulting expression can be simplified like in Section \ref{sec:master counting} and one finds
	\begin{equation}
		\label{eq:counting formula charged sectors}
		\dim \mathcal{H}_{\{q\}} =\tr P_{\{q\}} =\sum_{C} \pqty{\frac{\abs{G}}{\abs{C}}}^{E-V} \prod_{x \in \mathrm{sites}} (\dim q_x) \chi_{q_x}(C)^* \ .
	\end{equation}
	This expression is closely related to eq.\eqref{eq:master formula different sites}, which counts the number of singlets in the tensor product of the gauge Hilbert space with an external matter Hilbert space. The two expressions are related by the fact that the number of singlets in the tensor product $q_1 \otimes q_2 \otimes \cdots \otimes q_n$ is the same as the multiplicity of $q_n^*$ (the dual representation) in the tensor product $q_1 \otimes q_2 \otimes \cdots \otimes q_{n-1}$, but one requires $k= \dim q_n$ states to represent a $k$-plet, while only one state is required to represent a singlet.

	Finally, note that summing over all irreps $q$ one has the identity
	\begin{equation}
		\sum_q (\dim q) \chi_q(g) = \abs{G} \delta_{g,1} \ .
	\end{equation}
	This is just the character of the regular representation. Then summing eq.\eqref{eq:counting formula charged sectors} over all possible charged sectors one finds
	\begin{equation}
		\sum_{\{q\}} \dim \mathcal{H}_{\{q\}} = \abs{G}^E = \dim \Htot \ ,
	\end{equation}
	as expected.
	
	\subsection{Twisted boundary conditions}\label{sec:twisted}
	
	\begin{figure}
		\centering
		\begin{tikzpicture}
			\def\Lx{6};
			\def\Ly{5};
			\def\Lband{2};
			\draw[thick, line width=5pt, color=yellow] (\Lband,0) -- (\Lband,\Ly);
			
			\draw[step=1.0,black] (0,0) grid (\Lx,\Ly);
			
			\foreach \x in {0, ..., \Lx}
			{
				\draw[dashed, color=black] (\x,\Ly) -- (\x,\Ly+1);
			}
			
			\foreach \y in {0, ..., \Ly}
			{
				\draw[dashed, color=black] (\Lx,\y) -- (\Lx+1,\y);
			}
			
			\node[below left] at (\Lband,0) {$y$};
			\fill (\Lband,0) circle(0.10);
			\node[above left] at (\Lband,\Ly) {$x$};
			\fill (\Lband,\Ly) circle(0.10);
			\node[above left] at (\Lband,\Ly+1) {$y$};
			\fill (\Lband,\Ly+1) circle(0.10);
			\node[right] at (\Lband, \Ly+0.5) {$l$};
			
		\end{tikzpicture}
		\caption{A two dimensional periodic square lattice. The dashed links are the boundary links. The labelled link $l$ connects sites $x$ and $y$, but from the point of view of $l$, site $y$ belongs to a periodic copy. Removing all boundary links keeps the lattice connected: for example, the yellow path connects $x$ and $y$ without crossing the boundary.}
		\label{fig:periodic square lattice}
	\end{figure}
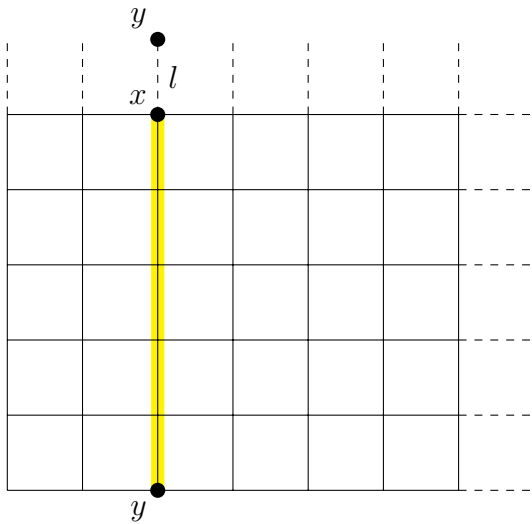
	
	The general formula eq.\eqref{eq:master counting formula} was derived under the assumption that the action of the Gauss law takes the same form on all links. This covers arbitrarily shaped lattices such as hypercubic, triangular, honeycomb, etc. with periodic, open or mixed boundary conditions. However, in some situations \textit{twisted} boundary conditions are useful. These boundary conditions are such that when crossing a boundary, a certain operation must be performed on the fields.
	
	Consider the prototypical example of a periodic square lattice, as shown in Fig. \ref{fig:periodic square lattice}. First of all, note that all degrees of freedom appearing in the Hilbert space belong (by definition) to the original lattice, rather than to the periodic copy. Therefore the total Hilbert space is the same for twisted and untwisted boundary conditions. Similarly, all on-site operators necessarily only act on degrees of freedom of the original lattice, and therefore do not feel the twisted boundaries. On the other hand, extended operators whose action extends over more than one site need to take into account the boundary twist. These include of course the Hamiltonian, but for our purposes, most importantly the Gauss law. In particular, this matters whenever a link lies on the boundary (these are the dotted links in Fig.\ref{fig:periodic square lattice}), i.e. it connects a site on the original lattice with one on a periodic copy. 
	
	Since gauge transformations act fully on-site for matter fields, we focus on the gauge sector. More concretely, we assume that the twisted boundaries are implemented by a function $\varphi: G \to G$. For the purposes of the Gauss law, this means that if $l=\expval{xy}$ is a boundary link, as shown in Fig.\ref{fig:periodic square lattice}, then from the point of view of the gauge transformation on link $l$, the group element associated to $y$ is $\varphi(g_y)$ rather than $g_y$, i.e. the transformation law is 
	\begin{equation}
		g_l \to g_x g_l \varphi(g_y)^{-1} \ .
	\end{equation}
	As discussed in Section \ref{sec:master counting}, a key assumption of our method for counting gauge invariant states is that gauge transformations form a group, i.e. $\mathcal{G}(\{g_x\}) \mathcal{G}(\{g'_x\})=\mathcal{G}(\{g_x g'_x\})$. This requirement is satisfied as long as $\varphi$ is a group homomorphism, i.e. $\varphi(g_1 g_2)=\varphi(g_1)\varphi(g_2)$. We will restrict ourselves to this case in the following, even though it does not cover all interesting boundary conditions. Examples include
	\begin{enumerate}
		\item \textit{No gauge transformation on the copy}. One takes $\varphi(g)=1$, which corresponds to not performing the gauge transformations on the other side of the boundary.
		\item \textit{C-periodic boundaries}. In this case, one performs a charge conjugation twist when crossing a boundary. We will discuss charge conjugation in more detail in Section \ref{sec:charge conjugation}. It turns out that it is implemented by a group automorphism $\varphi: G\to G$ (i.e. a bijective group homomorphism).
		\item \textit{Other symmetries implemented by automorphisms}. Similarly to charge conjugation, other symmetries are implemented by group automorphisms (this is discussed in Section \ref{sec:charge conjugation}). 
	\end{enumerate}
	We now discuss how to compute the number of gauge-invariant states in the case of boundary conditions twisted by a group homomorphism. Suppose that $l=\expval{xy}$ is a boundary link, like in Fig.\ref{fig:periodic square lattice}. This means that $x,y$ are both lattice sites, but that from the point of view of $l$ one of the sites (say $y$, like in the picture) lies \say{beyond the boundary}. This implies that, when computing the gauge transformation on link $l$, the group element associated to $y$ is $\varphi(g_y)$ rather than $g_y$. In other words, the transformation law is
	\begin{equation}
		\label{eq:twisted gauge transformation}
		\ket{g_l} \to \ket{g_x\, g_l\, \varphi(g_y)^{-1}} \ .
	\end{equation}
	What are the implications for the calculation of the number of gauge-invariant states? The calculation would proceed much in the same way as in Section \ref{sec:master counting}. One has again
	\begin{equation}
		\dim \Hphys = \tr P = \frac{1}{\abs{G}^V} \sum_{\{g_x\}} \tr \mathcal{G}(\{g_x\}) \ .
	\end{equation}
	The key takeaway from that section is that the trace of a gauge transformation $\mathcal{G}(\{g_x\})$ is non-zero only if for each link $l=\expval{xy}$, the elements $g_x$ and $g_y$ belong to the same conjugacy class. In the case of twisted boundaries, this implies that if $l=\expval{xy}$ is a boundary link, then $g_x$ and $\varphi(g_y)$ must belong to the same conjugacy class. Now assume that the lattice remains connected when all the boundary links are removed. This is generally satisfied except in peculiar cases and is again shown in Fig. \ref{fig:periodic square lattice}. Then there is also a path in the lattice connecting $x$ and $y$ without crossing a boundary. Therefore this means that $g_x$ and $g_y$ must also lie in the same conjugacy class in order for the trace to be non-zero. Thus $g_x$, $g_y$ and $\varphi(g_y)$ must all lie in the same conjugacy class. For notation, let $\partial X$ be the set of lattice sites which are endpoints of a boundary link. Then the trace of the projector yields a variant of eq.\eqref{eq:intermediate Hphys calculation}, i.e.
	\begin{equation}
		\dim \Hphys = \frac{1}{\abs{G}^V} \sum_C \sum_{\{g_x\} \in C} \pqty{\frac{\abs{G}}{\abs{C}}}^E \chi_U(C)^V \prod_{x \in \partial X} \delta(\varphi(g_x)\in C)\ ,
	\end{equation}
	where the delta function is $1$ if the condition is satisfied and zero otherwise. This is the same formula as eq.\eqref{eq:intermediate Hphys calculation}, the difference being that we must now enforce that $\varphi(g_y)$ also lies in the same conjugacy class as all the other elements. Now, to proceed, the sum over $\{g_x\}$ just gives a factor of $\abs{C}$ for each $x \not\in \partial X$ and otherwise can be placed inside the product on the right. One then obtains the final result eq.\eqref{eq:master formula twisted non homomorphism}. We are now ready to state the general counting formula for twisted boundary conditions.
	
	Let $G$ be an arbitrary finite gauge group. On an arbitrary lattice with $V$ sites and $E$ edges, with gauge transformations given by eq.\eqref{eq:gauge transformation generic} except on $\abs{\partial X}$ boundary sites where they are twisted by a group homomorphism $\varphi$ as in eq.\eqref{eq:twisted gauge transformation}, if the matter field representation has character $\chi_U$, then the number of gauge-invariant states is given by
	\begin{equation}
		\label{eq:master formula twisted non homomorphism}
		\dim \Hphys =\sum_C \pqty{\frac{\abs{G}}{\abs{C}}}^{E-V} \chi_U(C)^V \alpha(C)^{\abs{\partial X}}\ , \qquad \alpha(C)\equiv \frac{1}{\abs{C}}\sum_{g  \in C} \delta(\varphi(g)\in C) \ .
	\end{equation}
	The formula simplifies if $\varphi$ is a group automorphism, in which case it is invertible and one ends up summing only over the subset of conjugacy classes fixed by $\varphi$:
	\begin{equation}
		\label{eq:master formula twisted}
		\dim \Hphys =  \sum_{C,\, \varphi(C)=C} \pqty{\frac{\abs{G}}{\abs{C}}}^{E-V} \chi_U(C)^V \ .
	\end{equation}
	Note that since $\varphi$ is an automorphism, at least the identity in $G$ (which forms a singlet conjugacy class) satisfies $\varphi(1)=1$. Interestingly, in this case the formula does not depend in any way on the number of boundary links; it is unchanged if the boundary twist is imposed even on just one link. In the case of $C$-periodic boundary conditions, as explained in Appendix \ref{sec:charge conjugation} some groups admit an unambiguous definition of charge conjugation, in which case $\varphi(C)=C^{-1}$ (i.e. the conjugacy class of the group inverses). Thus one would sum only over those conjugacy classes which are self-inverse, $C=C^{-1}$.
	
	As highlighted in the introduction, this section illustrates how we are developing a \textit{toolkit} for the calculation of the number of gauge-invariant states in a wide variety of situations. We hope to have provided sufficient methods for the reader to perform the calculation in whatever situation is of their interest.
	
	\subsection{Charged states and twisted boundaries}
	
	Among all gauge transformations, some are \say{global} transformations, which act with the same group element $g_x=g$ on each site. We denote them as $Q(g)$ and, since they are gauge transformations, they again take the form eq.\eqref{eq:gauge transformation generic}:
	\begin{equation}
		\label{eq:global charge}
		Q(g) = \bigotimes_{l \in \mathrm{links}} L_{g} R_{g} \bigotimes_{x \in \mathrm{sites}} U_{g} \ .
	\end{equation}
	Note that we use the notation $Q$ even though it is a group object (i.e. it is unitary, not Hermitean). It should be emphasized that global gauge transformations are, nonetheless, gauge transformations. It is not possible to separate the \say{global part} of the gauge symmetry, since (because gauge transformations form a group) composing two non-global gauge transformation can very well be a global transformation. Thus all physical states satisfy $Q(g) \ket{\psi}=\ket{\psi}$, i.e. they are invariant under all $Q(g)$.
	
	Now we specialize the discussion to Abelian groups. In this case one has $L_g R_g = 1$, i.e. because group elements commute $g g' g^{-1}=g'$. This is just the statement that Abelian gauge fields do not carry charge. Thus the global transformations eq.\eqref{eq:global charge} do not transform the gauge fields, unless one imposes twisted boundary conditions as discussed in the previous section. In that case, one has instead $L_g R_{\varphi(g)}$ which is not necessarily the identity. Thus for Abelian gauge groups with twisted boundary conditions, one finds
	\begin{equation}
		\label{eq:twisted global charge}
		Q(g) = \bigotimes_{l \in \partial X} L_{g} R_{\varphi(g)} \bigotimes_{x \in \mathrm{sites}} U_{g} \ ,
	\end{equation}
	where $\partial X$ denotes the boundary links. It is then possible to split the global gauge transformation into a \say{bulk} part and a \say{boundary} part,
	\begin{equation}
		Q(g) = Q^{\mathrm{bulk}}(g) Q^{\mathrm{boundary}}(g) \ ,
	\end{equation}
	where
	\begin{align}
		Q^{\mathrm{bulk}}(g) &= \mathds{1} \otimes \bigotimes_{x \in \mathrm{sites}} U_{g} \label{eq:bulk charge}\ ,\\
		Q^{\mathrm{boundary}}(g) &=\bigotimes_{l \in \partial X} L_{g} R_{\varphi(g)} \otimes \mathds{1} \label{eq:boundary charge} \ .
	\end{align}
	Their interpretation is clear: the bulk charge $Q^{\mathrm{bulk}}$ is carried by the matter fields (as Abelian gauge fields are charge neutral), while the boundary charge $Q^{\mathrm{boundary}}$ is carried by the gauge field only, and it compensates exactly the bulk charge. In fact since their product is a gauge transformation, on physical states one has
	\begin{equation}
		Q^{\mathrm{bulk}}(g) Q^{\mathrm{boundary}}(g) \ket{\psi}=\ket{\psi} \ .
	\end{equation}
	Both $Q^{\mathrm{bulk}}$ and $Q^{\mathrm{boundary}}$ are gauge-invariant in the Abelian case, and therefore their action on physical states is well-defined. Moreover, neither of them is in general a gauge transformation and therefore they may act non-trivially on physical states. The bulk charge may therefore be used to identify charged configurations. It is in fact useful to establish when exactly $Q^{\mathrm{bulk}}$ (and therefore also $Q^{\mathrm{boundary}}$) is a gauge transformation (and therefore acts trivially on all physical states). This happens precisely when it takes the form eq.\eqref{eq:twisted global charge}, i.e. when $\varphi(g)=g$. We will make use of this fact later.
	
	This discussion is another way to formulate the well-known fact that twisted boundary conditions allow charged states which would otherwise be forbidden. It gives an explicit picture whereby the excess bulk charge is carried away by the gauge field through the twisted boundary. Using the formulas for the number of gauge-invariant states developed in the previous sections, we can quantify the presence of charged states. For example, we will be able to prove the famous fact that \say{there are no charged states on the torus}. Using this formalism, one can deduce what kind of charged states are present with a given geometry and boundary conditions.
	
	We focus on the Abelian group $\Z_N$. It is generated by a single element $\xi$ which satisfies $\xi^N=1$. It has $N$ irreducible representations indexed by their charge $q=0,1,\ldots, N-1$, i.e. they are given by $\rho_q(\xi^k) = \exp{\pqty{\frac{2\pi i q k}{N}}}$. The matter representation $U$ can in general be written as a direct sum of the irreducible representations $\rho_q$. So for simplicity we consider the situation where we assign a charge $q_x$ (i.e. the representation $\rho_{q_x}$) to each site $x$ (splitting $U$ into irreducibles, these results can be combined to form the general case). Then the total $\Z_N$ charge is given by
	\begin{equation}
		Q = \sum_x q_x \,\,(\mathrm{mod}\,\,N) \ .
	\end{equation}
	Then one has 
	\begin{equation}
		Q^{\mathrm{bulk}}(\xi^k) =  \exp{\pqty{\frac{2\pi i Q k}{N}}} \ .
	\end{equation}
	
	First of all let's consider the case where there's no twist. This includes most importantly periodic boundaries, but also open boundaries or mixed open/periodic boundaries (as long as the Gauss law is imposed everywhere in the same form). Then the boundary charge eq.\eqref{eq:boundary charge} is always trivial and the bulk charge is just a gauge transformation: thus bulk charge-neutrality follows by the Gauss law. Let's see this explicitly. Since $\Z_N$ is Abelian, all of its conjugacy classes are singlets and from the formula for the number of gauge-invariant states eq.\eqref{eq:master formula different sites} one has
	\begin{equation}
		\dim \Hphys = N^{E-V}  \sum_{k=0}^{N-1} \prod_{x \in \mathrm{sites}} e^{2\pi i q_x k/N} =  N^{E-V}  \sum_{k=0}^{N-1} e^{2\pi i Q k/N}  = \begin{cases}
			N^{E-V+1} & Q =
			0\\ 0 & Q \neq 0 \end{cases}\ .
	\end{equation}
	Thus indeed there are no gauge-invariant states with total non-zero charge. Another typical situation where this formula applies is that of dimer models, where the charge alternates $q_x = (-1)^x q$ on even/odd lattice sites and we have an equal number of each; then the total charge is zero. From the above, we know that the number of gauge-invariant states in fact equals that in the pure gauge theory.
	
	\begin{figure}
		\centering
		\begin{tikzpicture}
			\def\Lx{4};
			\def\Ly{3};
			
			\draw[step=1.0,black] (0,0) grid (\Lx,\Ly);
			
			\foreach \x in {0, ..., \Lx}
			{
				\draw[color=black] (\x,\Ly) -- (\x,\Ly+1);
			}
			
			\foreach \x in {0, ..., \Lx}
			\foreach \y in {0, ..., \Ly}
			{
				\fill (\x,\y) circle(0.10);
			} 
			
			\foreach \x in {0, ..., \Lx}
			{
				\fill[color=white] (\x,\Ly+1) circle(0.10);
				\draw[thick] (\x,\Ly+1) circle(0.10);
			}
			
		\end{tikzpicture}
		\caption{A two dimensional square lattice with open boundaries. Some links are \protect\say{dangling}, i.e. they are connected to the rest of the lattice via only one site.  Every link carries a gauge field, but on dangling links it is not dynamical. The matter field is present on all black sites, but not on white sites. Gauge transformations are imposed on black sites, but not on white sites. This figure is a two-dimensional version of Fig.2 in \protect\cite{Chung:2024hsq}.}
		\label{fig:open vacuum boundary}
	\end{figure}
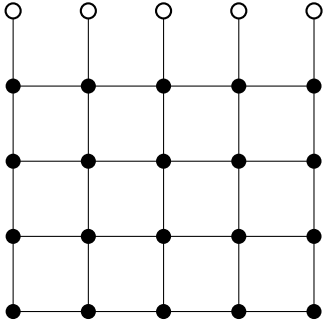
	
	One geometry which allows charged matter in the case of open boundaries consists in adding links with non-dynamical gauge fields which allow the charge to \say{escape}. An example is depicted in Fig.\ref{fig:open vacuum boundary}. Such boundary conditions have been recently considered in several works precisely for this purpose \cite{Verresen:2022mcr,Thorngren:2023ple,Chung:2024hsq}. This case can be treated as a sort of twisted boundary, where $\varphi(g)=1$. Since this is not an automorphism, we employ eq.\eqref{eq:master formula twisted non homomorphism}. In this case $\alpha(C)$ is non-zero only on the identity, and we find
	\begin{equation}
		\dim \Hphys = N^{E-V}  \ .
	\end{equation}
	which is non-zero irrespectively of the total charge $Q$. Note that in this case the bulk charge is conserved, since the extra gauge links are non-dynamical and therefore do not appear in the Hamiltonian.
	
	If one instead wants to maintain both locality and translation invariance while working with charged states, one solution is to use $C$-periodic boundary conditions \cite{PolleyWiese, KronfeldWiese, Wiese1992}, which have found applications in both particle \cite{Lucini:2015hfa} and condensed matter physics \cite{hornung2021mass, Mariani:2023yre, InfiniteZ}. With these boundary conditions, one performs a charge conjugation transformation on the fields when crossing a boundary. There are some subtleties in the definition of charge conjugation for arbitrary finite gauge groups, which we will discuss in the next section. As we will see there, for $\Z_N$, charge conjugation is implemented by group inversion, $g \to \varphi(g)= g^{-1}$.
	
	In this case, there's a difference for even and odd $N$. For odd $N$ there's no non-trivial element in $\Z_N$ which squares to the identity (this is similar to theories with gauge group $\Z$); on the other hand, for $N$ even, the element $g=\xi^{N/2}$ satisfies $g^2=1$, i.e. $\varphi(g)=g$ (this is similar to theories with gauge group $\U(1)$). Bulk transformations with $g=\xi^{N/2}$ are thus gauge transformations. This also implies that the bulk charge is only defined modulo two.
	
	Note that in this case the bulk charge does not commute with the Hamiltonian (unless it is a gauge transformation). Nonetheless we can check whether states with given overall bulk charge exist. According to the general formula for twisted boundaries eq.\eqref{eq:master formula twisted}, we sum only over those group elements which are self-inverse. For $N$ odd this is just the identity, and thus one finds
	\begin{equation}
		\dim \Hphys = N^{E-V} \ ,
	\end{equation}
	which is always non-zero, so charged states are always allowed with arbitrary total charge $Q$. For $N$ even, one gets instead
	\begin{equation}
		\label{eq:charged states ZN even}
		\dim \Hphys = N^{E-V}  \sum_{k=0, N/2} e^{2\pi i Q k/N} = N^{E-V} (1 + (-1)^{Q}) \ ,
	\end{equation}
	so some charged states are allowed (i.e. those with overall even charge). With these boundary conditions one can then create a state with overall non-zero bulk charge, with the gauge field carrying away the remaining flux through the boundary; in the infinite volume limit, one has then obtained a charged state (see also \cite{Fredenhagen:1983sn}).
	
	\subsection{Physical symmetries and charge conjugation}\label{sec:charge conjugation}
	
	In this section we discuss the implementation of certain symmetries in the gauge sector, in particular charge conjugation. The discussion applies to finite or compact Lie gauge groups. In $\SU(N)$ gauge theories, charge conjugation acts on non-Abelian gauge fields like complex conjugation \cite{Bietenholz_Wiese_2025}, i.e. $U_\mu(x) \to U_\mu(x)^*$. As we will see, this definition is however somewhat simplistic and it is only well-defined because of certain group-theoretical properties of $\SU(N)$ which do not necessarily apply in general.
	
	Let us examine the issues with this definition in more detail. We consider first the instructive case of Abelian groups, where this problem does not occur. For Abelian groups, all irreps are one-dimensional and therefore for any irrep $\rho$ one has
	\begin{equation}
		\rho(g^{-1})\equiv\rho(g)^\dagger =\rho(g)^* \ ,
	\end{equation}
	where we used the fact that representations can in general be chosen to be unitary. Therefore in this case there is an operation, $g \to g^{-1}$, which acts on abstract group elements, and complex conjugates all irreps. This is then the group-theoretical definition of charge conjugation for Abelian groups. Since all representations can be broken up into a direct sum of irreps, then group inversion complex conjugates all representations (up to a basis change).
	
	The two key ingredients in the above construction are first the fact that charge conjugation is implemented as an operation on abstract group elements, i.e. a map $G \to G$. Secondly, it is defined such that it complex conjugates \textit{all} representations at the same time. These two conditions ensure that it is well-defined and unambiguous. Now let us examine more in detail what happens for non-Abelian groups. In this case, group inversion only gives Hermitean conjugation, which is not what we want. Suppose that we take a faithful group representation in a specific basis and say that charge conjugation is implemented by complex conjugating these matrices, i.e. we identify the group elements with the matrices in this faithful representation. The first problem with this definition is that there's no guarantee that this operation would then also complex conjugate the representation matrices in other representations; the fact that this happens for $\SU(N)$ is quite remarkable. The second issue is that it appears to depend on which basis one chooses for the representation matrices. A striking example is the following: a group representation is said to be \textit{real} if there is a basis where all of its matrix elements are real-valued. If by charge conjugation we mean \say{complex conjugate the representation matrix}, then in the basis where the elements are real, charge conjugation would do nothing; in a different basis, instead, it would act non-trivially. In the rest of this section, we will discuss how these issues can be remedied.
	
	Before turning specifically to charge conjugation, we briefly discuss symmetries of the gauge sector implemented by group automorphisms $\tau: G \to G$, i.e. bijective group homomorphisms. From the above discussion, charge conjugation should be implemented by such a map. On the Hilbert space, such a symmetry acts via a unitary operator $U_\tau$ defined as
	\begin{equation}
		U_\tau \ket{g} \equiv \ket{\tau(g)} \ .
	\end{equation}
	For the unitarity of $U_\tau$ it is essential that $\tau$ is an automorphism. Since this is a gauge theory, we are interested in the action of symmetries which are not gauge transformations, as these would act trivially on physical states. The automorphisms of a group $G$ form a group $\mathrm{Aut}(G)$ (under composition). The automorphisms $\tau$ of the form $\tau(g) = g_* gg_*^{-1}$ for fixed $g_*$ are known as \textit{inner} and form a normal subgroup $\mathrm{Inn}(G) \leq \mathrm{Aut}(G)$. On the Hilbert space, an inner automorphism acts as 
	\begin{equation}
		U_\tau \ket{g} \equiv \ket{g_* gg_*^{-1}} \ ,
	\end{equation}
	which is the same as the action of a global gauge transformation. Thus symmetries implemented by an inner automorphism \say{do nothing} on $\Hphys$. So for many purposes one can simply quotient out the inner automorphisms, obtaining the so-called group $\mathrm{Out}(G)$ of \textit{outer} automorphisms,
	\begin{equation}
		\mathrm{Out}(G) \equiv \frac{\mathrm{Aut}(G)}{\mathrm{Inn}(G)} \ .
	\end{equation}
	We will keep this in mind in what follows. It is easy to check that a certain automorphism $\tau$ is a symmetry of the gauge Hamiltonian eq.\eqref{eq:gauge hamiltonian} if and only if it satisfies the following two conditions:
	\begin{enumerate}
		\item It preserves $h_B$, i.e. $h_B(\tau(g))=h_B(g)$.
		\item It preserves the subset $\Gamma$ of eq.\eqref{eq:group Laplacian}, i.e. $\tau(\Gamma)=\Gamma$. \footnote{An interesting consequence of this is that if a group $G$ has a non-inner automorphism $\tau$ which nonetheless preserves the conjugacy classes then it is a symmetry for all Hamiltonians of the form eq.\eqref{eq:gauge hamiltonian}, independent of the details. Such automorphisms have implications for the physical Hilbert space which were extensively discussed in \cite{Mariani:2024osg}.}
	\end{enumerate}
	We note in passing that outer automorphisms have been used to represent $CP$ symmetries in the context of finite flavour groups \cite{Chen:2014tpa, Trautner:2017vlz}. It does not appear, however, that there is a relation to the discussion in the present section.
	
	Now we return more specifically to charge conjugation. It should be implemented by an automorphism $\tau$ which should complex conjugate the representation matrices. It turns out that this is too restrictive and instead we simply require that for a representation $\rho$, then $\rho\circ \tau$ and $\rho^*$ are merely \textit{equivalent}, i.e. they are allowed to differ by a change of basis. Ideally, we would like this to be true for all representations $\rho$. To remove the equivalence condition, we can then take the trace (which is basis invariant). Note that two representations are equivalent if and only if they have the same character. If $\rho$ has character $\chi$, then equivalence of $\rho\circ \tau$ and $\rho^*$ is the same as requiring that they have the same character, i.e.
	\begin{equation}
		\label{eq:class-inversion}
		\chi(\tau(g))=\chi(g)^*=\chi(g^{-1}) \ .
	\end{equation}
	If $\tau$ satisfies eq.\eqref{eq:class-inversion} for all (irreducible) characters $\chi$ (and therefore for all representations) then it is called \textit{class-inverting}. Naturally, we would also expect that charge conjugation is a $\Z_2$ symmetry, i.e. $\tau(\tau(g))=g$. Such an automorphism is called \textit{involutory}. This motivates the following definition (see also \cite{Okubo}):
	
	\begin{definition}\label{def:charge conjugation transf}
		Given a gauge group $G$, a charge conjugation transformation is an involutory class-inverting group automorphism $\tau$. 
	\end{definition}
	
	If $\tau$ does not satisfy all requirements of Definition \ref{def:charge conjugation transf} but it inverts some characters (as in eq.\eqref{eq:class-inversion}) then we say that $\tau$ is \say{$C$-like}. We have the following definitions \cite{Sharp}:
	\begin{enumerate}
		\item A group $G$ admitting an involutory, class-inverting automorphism is called \textit{quasi-ambivalent}.
		\item A group $G$ admitting an involutory, class-inverting \textit{inner} automorphism (including the identity automorphism) is called \textit{ambivalent}. This is equivalent to all characters being real-valued.
	\end{enumerate}
	
	For ambivalent groups, charge conjugation is just a gauge transformation. For Lie groups commonly considered in lattice gauge theory, there is no ambiguity in the definition of charge conjugation, because of the following result:
	\begin{theorem} \cite{Sharp} All compact connected simple Lie groups (with the exception of $E_6$) are quasi-ambivalent. More precisely:
		\begin{enumerate}
			\item $\SU(N)$ for $N\geq 3$ as well as $\SO(4k+2)$, $\U(N)$ and $\GL(N)$ are quasi-ambivalent, but not ambivalent.
			\item $\SU(2)$, $\SO(2n+1)$, $\SO(4k)$ as well as $E_7, E_8, F_4, G_2$ are ambivalent. 
		\end{enumerate}
	\end{theorem}
	
	Thus for the most commonly considered case of $\SU(N)$, there is no ambiguity in the definition of charge conjugation. In fact for $N \geq 3$, $\SU(N)$ admits a unique outer automorphism which is precisely charge conjugation. On the other hand, $\SU(2)$ does not have non-trivial outer automorphisms, but charge conjugation is implemented by an inner automorphism (i.e. conjugation by $i\sigma_2$ in the fundamental representation), that is, it is given by a gauge transformation.  
	
	For finite groups, not many general results are known. While ambivalence can be easily checked by looking at the character table, no simple characterization is known for quasi-ambivalence. For example, it is known that finite quasi-ambivalent groups have even order \cite{Sharp}. For a generic finite gauge group $G$, we have the following situation:
	\begin{enumerate}
		\item $G$ is quasi-ambivalent, and there is a unique  involutory class-inverting outer automorphism. Then the situation is like in $\SU(N)$: there is an unambiguous definition of charge conjugation.
		\item $G$ is quasi-ambivalent, but there is more than one involutory class-inverting outer automorphism. In this case, multiple definitions of charge conjugation are possible.
		\item $G$ is not quasi-ambivalent. In this case, it may or may not be possible to define a charge conjugation operation, but it will fail to satisfy one or more of the conditions of Definition \ref{def:charge conjugation transf}. One would then look for an automorphism which complex conjugates the representation which effectively appears in the Hamiltonian. But this transformation might not be $\Z_2$ or it might depend on the chosen representation(s) in the Hamiltonian. 
	\end{enumerate}
	
	In particle physics, one is mostly interested in subgroups of $\U(1)$, $\SU(2)$ and $\SU(3)$. Note that quasi-ambivalence is not automatically transferred to subgroups. In fact, restricting an automorphism to a subgroup does not necessarily give an automorphism of the subgroup.
	
	The cyclic groups $\Z_N$, being Abelian, are quasi-ambivalent but not ambivalent (the automorphism being inversion, $g \to g^{-1}$). The dihedral groups $D_N$ are ambivalent (this is because the reflections have order $2$ and the rotations are conjugate to their inverse via the generator of reflections). Since the quaternions $Q$ share the same character table as $D_4$, they are also ambivalent. The non-Abelian group of order $21$, on the other hand, is not quasi-ambivalent and neither are the alternating groups $A_n$ for $n \geq 16$; the smallest non-quasi-ambivalent group has order $20$ \cite{Sharp}.
	
	\begin{table}[t]
		\centering
		\renewcommand{\arraystretch}{1.1}
		\setlength{\tabcolsep}{3pt}
		\begin{tabular}{lcccccc}
			\toprule
			\hspace{5em}$G$ & Order & \texttt{GAP} & Center & $\mathrm{Out}(G)$ & Amb.? & Quasi-amb.? \\
			\midrule
			$\SU(2)$
			& / & / & $\Z_2$ & triv. & yes & yes  \\[5pt]
			Binary tetrahedral
			& $24$ & \texttt{SL(2,3)} & $\Z_2$ & $\Z_2$ & no & yes  \\[5pt]
			Binary octahedral
			& $48$ & \texttt{SmallGroup(48,28)} & $\Z_2$ & $\Z_2$ & yes & yes \\[5pt]
			Binary icosahedral
			& $120$ & \texttt{SL(2,5)} & $\Z_2$ & $\Z_2$ & yes & yes \\
			\midrule
			$\SU(3)$
			& / & / & $\Z_3$ & $\Z_2$ & no & yes  \\[5pt]
			$\Sigma(108)$
			& $108$ & \texttt{SmallGroup(108,15)} & $\Z_3$ & $\Z_2 \times \Z_2$ & no & yes  \\[5pt]
			$\Sigma(216)$
			& $216$ & \texttt{SmallGroup(216,88)} & $\Z_3$ & $S_3$ & no & no \\[5pt]
			$\Sigma(648)$
			& $648$ & \texttt{SmallGroup(648,532)} & $\Z_3$ & $\Z_6$ & no & yes \\[5pt]
			$\Sigma(1080)$
			& $1080$ & \texttt{SmallGroup(1080,260)} & $\Z_3$ & $\Z_2 \times \Z_2$ & no & no \\
			\bottomrule
		\end{tabular}
		\caption{Summary of some group theoretical properties of crystal-like subgroups of $\SU(2)$ and $\SU(3)$. The columns are group $G$, its order, (one of) its  \texttt{GAP} names, its center, its outer automorphism group, whether the group is ambivalent, and whether it is quasi-ambivalent. For some groups, further description of their automorphisms is given in the main text.}
		\label{tab:subgroup summary}
	\end{table}
	
	Now we consider commonly studied crystal-like subgroups of $\SU(2)$ and $\SU(3)$. These groups can be studied using the software package \texttt{GAP} \cite{GAP4}. Useful references include \cite{Golasinski, Ludl:2009ft, Ludl:2011gn, Flyvbjerg:1985ad, Coquereaux:2012at, Assi:2024pdn}. A summary of relevant properties of these groups can be found in Table \ref{tab:subgroup summary}.
	
	First of all, we discuss the three crystal-like subgroups of $\SU(2)$. We have the following result:
	
	\begin{proposition}
		The binary octahedral and binary icosahedral group are ambivalent. The binary tetrahedral group is quasi-ambivalent, but not ambivalent.
	\end{proposition}
	
	Ambivalence is easily checked by looking at the character table. Quasi-ambivalence needs to be checked explicitly for the binary tetrahedral group. In fact it has twelve class-inverting automorphisms, six of which are involutory. The unique non-trivial outer automorphism of the binary tetrahedral group is class-inverting.
	
	Now we discuss their symmetries arising from outer automorphisms. As we have just seen, for the binary tetrahedral group the unique non-trivial outer automorphism acts as charge conjugation. Unlike $\SU(2)$, this action is non-trivial on physical states (i.e. it is not a gauge transformation). With a Hamiltonian of the form eq.\eqref{eq:gauge hamiltonian}, it is always a symmetry. The other two groups have real-valued characters, like $\SU(2)$, but, unlike $\SU(2)$, they each admit a non-trivial outer automorphism which could potentially represent a symmetry. In both cases, it is reasonable to pick the representation in the gauge Hamiltonian as the restriction to the subgroup of the fundamental representation of $\SU(2)$ (in both cases, it is irreducible). Then for both groups the outer automorphism turns out not to be a symmetry. Finally, we note that none of these subgroups of $\SU(2)$ (or $\SU(3)$) admit class-preserving automorphisms. These are important, as explained in \cite{Mariani:2024osg}.
	
	Next we discuss the four crystal-like subgroups of $\SU(3)$:
	
	\begin{proposition}
		The crystal-like $\SU(3)$ subgroups $\Sigma(108)$ and $\Sigma(648)$ are quasi-ambivalent but not ambivalent. On the other hand, $\Sigma(216)$ and $\Sigma(1080)$ are not quasi-ambivalent.
	\end{proposition}
	
	Note that these groups are referred to by other symbols as well, for example $\Sigma(1080)$ is sometimes referred to as $S(1080)$, $\Sigma(360 \times 3)$ or $\Sigma(360 \phi)$ and similarly for the other groups. Ambivalence and quasi-ambivalence have been checked by hand using \texttt{GAP}.
	
	We will only discuss in more detail the automorphisms of $\Sigma(1080)$, since it is the most interesting. Unlike $\SU(3)$, it is not quasi-ambivalent, which means that charge conjugation is not defined unambiguously. Its outer automorphism group is $\Z_2 \times \Z_2$ and its elements can be thought of as acting by permuting the four three-dimensional irreducible representations of $\Sigma(1080)$. Precisely one outer automorphism exchanges the restrictions to $\Sigma(1080)$ of the fundamental and antifundamental representations of $\SU(3)$ (which are irreps of $\Sigma(1080)$), and therefore can be taken in practice to be charge conjugation. However, as anticipated, it is not class-inverting and therefore charge conjugation might be broken depending on which representations are included in the Hamiltonian or in the operators. This may be an issue especially in light of systematic improvement. If one picks the gauge Hamiltonian  eq.\eqref{eq:gauge hamiltonian} in the restriction of the fundamental of $\SU(3)$, then the other independent outer automorphism is not a symmetry.
	
	\section{Scalar fields}\label{sec:scalar}
	
	In this section, we instantiate the general construction of Section \ref{sec:general} to the case of scalar field theories with finite gauge symmetry. Here the key problem is that bosonic quantum field theories have a locally infinite-dimensional Hilbert space, and therefore a procedure for making this finite-dimensional is required. Several constructions have been proposed for this purpose in the literature \cite{Jordan:2011ci,Klco:2018zqz,Barata:2020jtq}. Here we take an agnostic approach and consider a rather general construction compatible with arbitrary finite gauge symmetry. We then discuss various examples, including how certain models encode a vast class of scalar field theories. We then give a formula for the number of gauge-invariant states.
	
	\subsection{Quantization of scalars}
	
	Suppose that the classical configuration space of the scalar field is $\mathcal{C}$. Then the quantum Hilbert space is the space of wavefunctions $\psi: \mathcal{C} \to \C$ which assign to each classical configuration its quantum probability amplitude. Therefore the quantum Hilbert space is $L^2(\mathcal{C}).$\footnote{Note that this is also the perspective arising from the transfer-matrix approach: one defines a Hilbert space of kets indexed by the classical configurations, i.e. those integrated over in the path-integral.} Scalar fields are associated to lattice sites, and therefore the classical configuration space $\mathcal{C}$ can be broken up as a product of local configuration spaces $S$, i.e.
	\begin{equation}
		\mathcal{C} = \bigtimes_{x \in \mathrm{sites}} S \ .
	\end{equation}
	Then the corresponding Hilbert space breaks up as a tensor product of the corresponding spaces,
	\begin{equation}
		L^2(\mathcal{C}) = \bigotimes_{x \in \mathrm{sites} } L^2(S) \ .
	\end{equation}
	For example, an $n$-component real scalar on a single-site lattice lives in the classical space $S=\R^n$. Then the quantum Hilbert space is the space of square-integrable functions $L^2(\R^n)$, which is infinite-dimensional.
	
	To obtain a finite-dimensional Hilbert space, the local configuration space $S$ must be finite. So we take $S$ to be an arbitrary finite set. Then square-integrability is always satisfied and the quantum Hilbert space is just the complex vector space $\C[S]$ of dimension $\abs{S}$. It has a privileged orthornormal basis of vectors $\{\ket{s}\}$ for $s \in S$. As is well-known it is not possible to implement canonical commutation relations in a finite-dimensional space \cite{Ercoetal1, Ercoetal2}. Thus in this section we do not make use of creation and annihilation operators. 
	
	For our purposes, the quantum Hilbert space of scalar matter is given by a tensor product of local Hilbert spaces on each site,
	\begin{equation}
		\label{eq:scalar Hilbert space}
		\mathcal{H}_S =\bigotimes_{x \in \mathrm{sites}} \C[S] \ .
	\end{equation}
	A prototypical example of this construction is the Ising model. In this case the scalar field locally takes values in $S=\Z_2 =\{1,-1\}$ and the local quantum Hilbert space is just $\C[\Z_2] \cong \C^2$. We will give more examples in just a moment.
	
	How does the gauge symmetry act on the Hilbert space? On each local space $S$, the symmetry is classically implemented generically by a \textit{group action}, i.e. a map $G \times S \to S$ which assigns to each element $s \in S$ its transformed element $g \cdot s$ by $g \in G$. Group actions are required to be such that the identity $1 \in G$ \say{does nothing}, i.e. $1 \cdot s = s$ and they must also respect group multiplication, i.e. $g_1 \cdot (g_2 \cdot s) \equiv (g_1 g_2) \cdot s$. Again picking the example of the Ising model, one has $S=\Z_2$ and a $\Z_2$ symmetry; in this case the group can be thought of as acting on itself via left multiplication. 
	
	\subsection{Examples}\label{sec:scalar examples}
	
	In this section, we give several examples of scalar fields valued in a set $S$ which carries an action of a group $G$. Note that typically the group action can be chosen to be faithful: if it isn't then one can mod out the kernel of the action from $G$ and obtain a new group which acts faithfully.
	
	For concreteness, in the Lagrangian formulation a typical action for a pure scalar field theory might take the form
	\begin{equation}
		\label{eq:pure scalar action}
		S = \sum_{\expval{xy}} K(s_x, s_y) + \sum_x V(s_x) \ ,
	\end{equation}
	where $s_x$ is an $S$-valued scalar field. For the theory to be invariant under a global symmetry $G$ given by a group action on $S$, both $V$ and $K$ must be $G$-invariant, i.e. $V(g \cdot s)=V(s)$ and $K(g \cdot s , g \cdot s')=K(s,s')$. We also assume that $K$ is symmetric in the two arguments.  Generally, we can gauge a subgroup $H < G$ by introducing an $H$-valued gauge field $h_l$ on lattice links.
	
	We briefly consider the consequences of gauging a subgroup on the global symmetry. In fact the global symmetry group must now act also on the gauge field, and its action must be compatible with the fact that $H$ is a subgroup; thus a global $G$ transformation given by $g \in G$ acts on the gauge field as $h_l \to g h_l g^{-1}$. It is easy to check that this is a symmetry, as long as the action is well-defined: i.e. $g h_l g^{-1}$ must again be an element of $H$. The set of elements of $G$ with this property forms a subgroup of $G$ known as the \textit{normalizer} $N_G(H)$ of $H$ in $G$ \cite{Serre}. It is defined as
	\begin{equation}
		N_G(H) = \{ g \in G | g H g^{-1} \subset H \} \ .
	\end{equation}
	In particular, $H$ is a normal subgroup of $N_G(H)$. All elements of $N_G(H)$ are thus global symmetries, but the elements which also lie in $H$ act as global gauge transformations (i.e. trivially on the physical Hilbert space $\Hphys$). They must therefore be modded out. Therefore when one gauges a subgroup $H$ of a global symmetry $G$, the residual global symmetry group is given by
	\begin{equation}
		N_G(H)/H  \ .
	\end{equation}
	Note that since $H$ is normal in $N_G(H)$, this is always a group.

	When introducing a gauge field, the action splits into a gauge part and a matter part, $S=S_G+S_M$ where $S_G$ is (say) the Wilson action and $S_M$ is the same as eq.\eqref{eq:pure scalar action} with a minimally coupled gauge field, i.e.
	\begin{equation}
		\label{eq:gauge scalar action}
		S_M = \sum_{\expval{xy}} K(s_x, h_{\expval{xy}} \cdot s_y) + \sum_x V(s_x) \ .
	\end{equation}
	Under mild assumptions on $K$, a straightforward transfer-matrix argument then gives the Hilbert space eqs.\eqref{eq:total Hilbert space},\eqref{eq:scalar Hilbert space} and a Hamiltonian of the form $H=H_G+H_M$ where $H_G$ is given in eq.\eqref{eq:gauge hamiltonian} and $H_M$ takes the form
	\begin{equation}
		H_M = \lambda \sum_x \hat{\Delta}_x + \sum_{\expval{xy}} K(s_{x}, h_{\expval{xy}} \cdot s_{y})  + \sum_{x} V(s_{x}) \ ,
	\end{equation}
	where $\hat{\Delta}$ acts on $\C[S]$ as a Laplacian in $S$-space:
	\begin{equation}
		\hat{\Delta} = \sum_{(r,s) \in \hat{\Gamma}} \ket{r}\bra{s} \ , \qquad \hat{\Gamma} = \{(r,s): \mathrm{min} K(r,s) \, | \, r,s \in S \, , r \neq s \} \ .
	\end{equation}
	As we will see, some models have $S=G$, a finite group, in which case $\hat{\Delta}$ is just the group Laplacian eq.\eqref{eq:group Laplacian}. Now we consider some more specific examples:
	
	\textit{Ising model}. As already discussed, the quantum Ising model has $S = \Z_2$ with a $\Z_2$ group action given simply by group multiplication.  In this case we take $K(s_1,s_2)=(1-s_1 s_2)$. Similarly, one can take $\Z_N$ valued scalar fields which carry an action of $\Z_N$ by group multiplication.
	
	\textit{Principal chiral field}. A generalization of the Ising model is the principal chiral field. In this case the scalar is valued in an arbitrary finite group, $S \cong G$. The symmetry group in this case is two copies of $G$, i.e. $G_L \times G_R$ which act by left and right multiplication, i.e.
	\begin{equation}
		g \to g_L g g_R^{-1} \ .
	\end{equation}
	Note that $G_L \times G_R$ does \textit{not} act faithfully on $G$; this is because if $g_L=g_R \in Z(G)$, the center of $G$, then $g_L g g_R^{-1} = g$ for all $g$. To obtain a faithful action, one needs to mod out the diagonal center subgroup, i.e. $G_L\times G_R/Z(G)$ acts faithfully on $G$ \footnote{This fact is familiar from the Standard Model \cite{Bietenholz_Wiese_2025}. The Higgs field carries an $\SU(2)_L \times \SU(2)_R$ action, but the Higgs sector only has a $(\SU(2)_L \times \SU(2)_R) / \Z_2 \cong O(4)$ symmetry, which is also apparent by decomposing the Higgs into four real scalars.}. For the kinetic term, one picks $K(g_1, g_2) = \Re\bqty{\chi(1)-\chi(g_1 g_2^{-1})}$ where $\chi$ is a faithful character of $G$. Note that if $G$ is Abelian, the left and right actions are the same and one only has a single $G$ symmetry, thus reducing to the Ising case. This model and its gauged version are especially interesting because, if $G$ is a compact Lie group, they describe effective theories such as chiral perturbation theory \cite{Gasser:1983yg} (also on the lattice \cite{Chandrasekharan:2003wy}). Recently, such effective theories have been considered for quantum simulation \cite{Watson:2023oov}.
	
	\textit{Coset models}. In these models, one has a group $G$ and a subgroup $H < G$. Then one takes $S=G/H$, the set of $H$ cosets, and $G$ acts on $G/H$ via left multiplication, i.e. $g \cdot g_1 H \equiv (gg_1)H$. These models are well-studied when $G$ is a Lie group. They include for example sigma models with target space $S^{n} \cong \Or(n+1)/\Or(n)$ and more generally Grassmannian models with target space $\mathrm{Gr}(r,n) \cong \Or(n)/(\Or(r) \times \Or(n-r))$. Also $\C \mathrm{P}^n$ models take this form, since $\C \mathrm{P}^n \cong \U(n+1)/(\U(n) \times \U(1))$. Similarly the Villain action for $\U(1)$-valued scalars is realized this way via $\U(1) \cong \R/\Z$ \cite{Villain:1974ir}. By replacing the Lie groups with an approximation as finite groups, one can construct scalar coset models with exact finite group symmetry.
	
	\textit{Models with transitive group actions}. A group action is said to be \textit{transitive} if given any two elements $s, s' \in S$ there is a $g \in G$ such that $s'=g\cdot s$. For example $\U(1)$ acts on $\C$ via multiplication but not transitively (because elements with different absolute value cannot be connected by the action), but if we restrict the action to $\U(1) \subset \C$ then the action is transitive. As discussed in Appendix \ref{sec:group actions}, every action of this form is equivalent to an action of $G$ on $G/H$ for some subgroup $H < G$. Therefore the \textit{coset models} above represent the general case of transitive group actions.
	
	\textit{Linear models}. Many scalar models of interest carry a linear action of a group, i.e. for example $S=\R^n$ carries a linear action of $\Or(n)$. A linear action is just a group representation. To obtain a finite-dimensional Hilbert space, $S$ must be finite. Linear spaces with finite $S$ take the form $S= (\Z_N)^n$. These are proper vector spaces only if $N$ is prime (so that $\Z_N=0,1,2,\ldots ,N-1$ is a field, i.e. all elements admit a multiplicative inverse), otherwise $\Z_N$ is only a ring and $S$ merely a \say{module}. But this does not matter much for our purposes. Thus we take $S= (\Z_N)^n$ and a linear action of $G$ on $S$ (group representations over these spaces have been studied by mathematicians \cite{Serre}). The quantum Hilbert space is then just $\C[(\Z_N)^n]$, i.e.\ basis states are indexed by vectors in $(\Z_N)^n$ (tuples of $n$ elements of $\Z_N$). One could then take for example $K = -\vec v_x \cdot \vec v_y$ for $\vec v \in (\Z_N)^n$. 
	
	\textit{Mixed models}. A complex scalar field $\Phi \in \C$ with a $\U(1)$ symmetry can be decomposed as $\Phi=R e^{i\theta}$. Here only the angular part is affected by the symmetry and the radial part is instead invariant. For example in the Standard Model, the Higgs can be written as $\Phi=R U$ where $R \geq 0$ and $U \in \SU(2)$ \cite{Bietenholz_Wiese_2025}. In the finite-dimensional setting, such models take the form $S = G \times \Sigma$ where $G$ is the \say{angular} part and $\Sigma$ the \say{radial} part. The symmetry group $G$ acts by left multiplication on the first factor. One could take $R \in \Sigma \cong Z_N=0,1,2, \ldots, N-1$ a kinetic term of the form $K=\tr(g_x^{-1} g_y) R_x R_y$ and a potential which depends on $R$ only.
	
	\textit{Models with free group actions}. A group action is said to be \textit{free} (as in \textit{fixed-point free}) if it has no fixed points, i.e. $g \cdot s=s$ for even one $s$ implies $g=1$. As discussed in Appendix \ref{sec:group actions}, all such group actions are equivalent to the action of $G$ on $G \times \Sigma$ where $\Sigma$ is an arbitrary set and $G$ acts on the first factor only, by left multiplication. Therefore if the action is free, it is immediately of the form of the mixed models above.
	
	\textit{Hardcore bosons}. One can also take spin operators which commute on different sites and anticommute on the same sites. The simplest such model is the spin-$\tfrac12$ model with $\SU(2)$ symmetry with kinetic term $\sim -\vec{\sigma}_x \cdot \vec{\sigma}_y$. The local Hilbert space is acted upon by the Pauli matrices and is thus isomorphic to $\C^2$. An arbitrary $\SO(3)$ rotation of the three Pauli matrices onto each other (i.e. $\sigma^i \to R^{ij} \sigma^j$) leaves the Hamiltonian invariant; via the $\SU(2) \to \SO(3)$ double cover, this is implemented as operators by a unitary $U$ such that $U \sigma^i U^\dagger = R_{ij} \sigma^j$ and therefore the states in the Hilbert space carry a representation of $\SU(2)$ under which they transform as $\ket{\psi} \to U \ket{\psi}$. Similar generalizations are possible for higher spin.
	
	\subsection{Counting gauge-invariant states}
	
	We are now ready to compute the number of gauge-invariant states. We have seen that the gauge group $G$ acts on $S$ via a certain group action. Then for each $g \in G$, one has an operator $U_g$ which acts on the Hilbert space $\C[S]$ via the classical group action. That is, given the preferred basis $\{\ket{s}\}$ of $\C[S]$ for $s \in S$, we set
	\begin{equation}
		\label{eq:Ug action}
		U_g \ket{s} \equiv \ket{g \cdot s} \ ,
	\end{equation}
	where $g \cdot s$ is the group action of $g \in G$ on $s \in S$. In other words, the symmetry acts on the quantum states via its classical action on a basis. It is then extended on the whole $\C[S]$ by linearity. It is not hard to show that $g \to U_g$ is a unitary group representation, i.e. it satisfies $U_{g_1} U_{g_2} = U_{g_1 g_2}$ and $U_{g^{-1}}=U_g^\dagger$.
	
	Thus the gauge group, which acts by an arbitrary group action on the classical configuration space $S$, acts via the group representation $U$ on $\C[S]$. As per the general construction of Section \ref{sec:master counting}, we need to compute the character (i.e. the trace) of $U$. Explicitly we have
	\begin{equation}
		\chi_U(g) = \sum_{s \in S} \bra{s} U_g \ket{s} = \sum_{s \in S} \braket{s | g \cdot s} = \sum_{s \in S} \delta(s, g\cdot s) \ .
	\end{equation}
	In other words, the character $\chi_U(g)$ is just given by the number of elements of $S$ fixed by $g \in G$. Calling this set $\mathrm{Fix}(g)$, we have simply
	\begin{equation}
		\chi_U(g) = \abs{\mathrm{Fix}(g)} \ .
	\end{equation}
	One can also show directly that indeed $\abs{\mathrm{Fix}(g)}$ only depends on the conjugacy class of $g$. We then have from the general counting formula eq.\eqref{sec:master counting} in the untwisted case, that the number of gauge-invariant states is given by
	\begin{equation}
		\label{eq:counting formula scalars}
		\dim \Hphys = \sum_{C} \pqty{\frac{\abs{G}}{\abs{C}}}^{E-V} \abs{\mathrm{Fix}(C)}^V \ .
	\end{equation}
	Note that the identity in $G$ always forms a singlet conjugacy class and fixes all elements of $S$, so $\abs{\mathrm{Fix(1)}}=\abs{S}$. Moreover, it is clear that the character is always non-negative; therefore the number of gauge-invariant states is never zero. For completeness, the size of the total Hilbert space is instead
	\begin{equation}
		\dim \Htot = \abs{G}^E \abs{S}^V \ .
	\end{equation}
	
	Many bases are possible for $\Hphys$. A simple option is to use the spin-network basis as in \cite{Baez, Burgio, MPE} by simply inserting a factor of the local matter Hilbert space in each local invariant subspace. Alternatively, the gauge field can be gauge-fixed on a maximal tree as in \cite{Mariani:2024osg, Grabowska:2024emw, Burbano:2024uvn}. On the other hand, one might want to remove the matter fields as much as possible. If the group action is transitive (as discussed in Section \ref{sec:scalar examples}) then the matter field can be gauge-fixed everywhere to a specific value (although one still needs to take appropriate superpositions to obtain gauge-invariant states, as explained in \cite{Mariani:2024osg}). More generally, other techniques have been devised to eliminate matter fields \cite{Zohar:2018cwb, Zohar:2019ygc}.

	\section{Fermion fields}\label{sec:fermions}
	
	In this section, we discuss the quantization of fermions and how the gauge group acts on the Hilbert space. We then compute the character of this representation and give a formula for the number of gauge-invariant states. This formula depends in practice on some peculiarities of the chosen fermionic formulation, so we discuss naive, Wilson and staggered fermions in more detail. Finally, we give a brief outline of how fermion parity superselection can be incorporated into the counting.
	
	\subsection{Quantization of fermions}
	
	Classically, fermion fields live in a vector space $V_F$. In a lattice field theory, the classical fermion space $V_F$ decomposes as a direct sum of local spaces on each site,
	\begin{equation}
		\label{eq:classical fermion vector space}
		V_F = \bigoplus_{x \in\mathrm{sites}} V_x \ .
	\end{equation}
	Each local space $V_x$ can be decomposed as a tensor product of spin (spacetime) and gauge degrees of freedom:
	\begin{equation}
		\label{eq:Vx decomposition}
		V_x \cong (V_x)_{\mathrm{spin}} \otimes (V_x)_{\mathrm{gauge}} \ .
	\end{equation}
	The spin degrees of freedom carry a representation of the Lorentz group, for example $(V_x)_{\mathrm{spin}}\cong \C^4$ for Dirac fermions in $3+1$ dimensions, or $(V_x)_{\mathrm{spin}}\cong \C$ for spinless fermions. On the other hand, the vector space $(V_x)_{\mathrm{gauge}}$ carries an arbitrary representation $\rho$ of the gauge group $G$. If one has multiple fermion flavours each transforming in a specific representation of $G$, then $V_x$ is the direct sum of subspaces of the form eq.\eqref{eq:Vx decomposition}. The gauge group $G$ acts non-trivially only on the second factor of $V_x$, i.e. it acts on each site as
	\begin{equation}
		\mathds{1} \otimes \rho \ .
	\end{equation}
	Since the gauge group acts non-trivially only on the gauge degrees of freedom, we will ignore spin and flavour indices for the moment and only restore them at the end.
	
	The quantization of fermion fields proceeds as follows. To each (orthonormal) basis vector $e_i \in V_F$ we associate a pair of creation and annihilation operators $\psi_i, \psi_i^\dagger$ for $i=1, \ldots , N$ where $N=\dim{V_F}$. Here $i$ is an arbitrary index. These operators are required to satisfy canonical anticommutation relations,\footnote{As noted by Wilson \cite{Wilson:1977nj}, it is also possible to quantize the fermions in an alternative way closely mirroring the quantization of bosons in Section \ref{sec:scalar}. That is, the fermion Hilbert space can be seen as the space of wavefunctions $f: V_F \to \C$, where however the argument of $f$ is taken to be a Grassmann variable. For example for a single spinless fermion, $V_F \cong \C$, and the quantum Hilbert space is that of functions $f(\psi^\dagger)$ of a single Grassmann variable $\psi^\dagger$ (with an appropriate inner product). Such a function can be expanded as $f(\psi^\dagger)=a+b\psi^\dagger$, giving the familiar two-level system. This approach leads to the same construction as in the main text.  }
	\begin{equation}
		\{ \psi_i, \psi_j\}=\{ \psi_i^\dagger, \psi_j^\dagger\} = 0 \ ,\qquad \{\psi_i, \psi_j^\dagger\} = \delta_{ij} \ .
	\end{equation}
	The quantum Hilbert space of fermions $\mathcal{H}_F$ is then the Fock space constructed from the creation and annihilation operators. Given a vacuum state $\ket{\Omega}$ which satisfies
	\begin{equation}
		\psi_i \ket{\Omega} = 0 \quad \forall i \ ,
	\end{equation}
	the Hilbert space is then constructed by acting with creation operators on the vacuum to form the states
	\begin{equation}
		\label{eq:typical fermionic state}
		\psi_{i_1}^\dagger \psi_{i_2}^\dagger \cdots \psi_{i_k}^\dagger \ket{\Omega} \ .
	\end{equation}
	Due to the anticommutation relations, these states are totally antisymmetric in the indices. Therefore an (orthonormal) basis of states for $\mathcal{H}_F$ is given by all states of the form \eqref{eq:typical fermionic state} with $i_1 < i_2 < \cdots < i_k$ for $k=0,1,\ldots, N$. A state with $k$ creation operators is referred to as having \textit{rank} $k$. The requirement that the indices be ordered is equivalent to the requirement that they be different. Thus the number of basis states of the rank $k$ subspace of $\mathcal{H}_F$ is given by the number of ways of choosing $k$ different indices out of $N$ possibilities, which is just $\binom{N}{k}$. Thus
	\begin{equation}
		\dim \mathcal{H_F} = \sum_{k=0}^N \binom{N}{k} = 2^N \ ,
	\end{equation}
	which follows from the binomial theorem. Now note that the subspace of rank $1$ states in $\mathcal{H}_F$ can be identified with $V_F$ via $e_i \leftrightarrow \psi_i^\dagger \ket{\Omega}$. Under this identification, $\mathcal{H}_F$ can be identified with the \textit{exterior algebra} (also known as the \textit{Grassmann algebra}) $\Lambda(V_F)$ of antisymmetric tensors on $V_F$, whereby
	\begin{equation}
		\label{eq:identification basis HF Lambda}
		\psi_{i_1}^\dagger \psi_{i_2}^\dagger \cdots \psi_{i_k}^\dagger \ket{\Omega} \qquad \leftrightarrow \qquad e_{i_1} \wedge e_{i_2} \wedge \cdots \wedge e_{i_k} \ ,
	\end{equation}
	where $\wedge$ is the wedge product. Note that $\Lambda(V_F)$ is an algebra under the wedge product, but here we are only interested in it as a vector space. It can be split as a direct sum of subspaces of rank $k$ tensors,
	\begin{equation}
		\Lambda(V_F) = \bigoplus_{k=0}^N \Lambda^k(V_F) \ ,
	\end{equation}
	where again $N=\dim V_F$. The exterior algebra moreover satisfies a property most useful for our purposes, i.e.\footnote{This is not directly relevant for our purposes, but note that to ensure the correct anticommutativity of tensors, the product of two pure tensors on the right-hand side must be interpreted under the so-called \say{Koszul sign rule} whereby $(a \otimes b) (c\otimes d) = (-1)^{\mathrm{deg}(b)\mathrm{deg}(c)} ac \otimes bd$.}
	\begin{equation}
		\label{eq:exterior algebra direct sum property}
		\Lambda(V_1 \oplus V_2 ) \cong \Lambda(V_1)\otimes \Lambda(V_2) \ .
	\end{equation}
	For a proof, see Chapter III Section 7.7 of \cite{Bourbaki1998-oi}. In any case, we will prove the corresponding statement for group representations after eq.\eqref{eq:exterior algebra representation factorization}, which is the result we need in practice. As an example application of eq.\eqref{eq:exterior algebra direct sum property} one has
	\begin{equation}
		\Lambda(\C^N) \cong \Lambda(\C)^{\otimes N} \cong (\C^2)^N \ ,
	\end{equation}
	which is just the familiar representation of the fermionic Fock space in the occupation number basis. Considering the subspace of rank $k$ on both sides of eq.\eqref{eq:exterior algebra direct sum property}, one obtains the equivalent property
	\begin{equation}
		\Lambda^k(V_1 \oplus V_2) \cong \bigoplus_{i=0}^k \Lambda^i V_1 \otimes \Lambda^{k-i} V_2 \ .
	\end{equation}
	For the classical fermion vector space $V_F$ in eq.\eqref{eq:classical fermion vector space}, the property \eqref{eq:exterior algebra direct sum property} translates into the statement
	\begin{equation}   
		\label{eq:fermion fock space site decomposition.}
		\mathcal{H}_F \cong \Lambda(V_F) = \Lambda\pqty{\bigoplus_{x \in\mathrm{sites}} V_x} \cong \bigotimes_{x \in\mathrm{sites}} \Lambda(V_x) \ .
	\end{equation}
	Thus the fermion Hilbert space can be split as a tensor product of local fermion Hilbert spaces. As familiar from the Jordan--Wigner decomposition, creation and annihilation operators, however, act non-locally. Note that the tensor product decomposition relies on an (arbitrary) implicit ordering of the lattice sites, but this will have no impact on what follows.
	
	\subsection{Gauge group representation}
	
	Gauge transformations act on $V_x$ by a representation $\rho$. What is the corresponding action on $\mathcal{H}_F \cong \Lambda(V_F)$? Suppressing for the moment spinor and flavour indices, creation and annihilation operators $\psi_a(x),\psi_a^\dagger(x)$ carry a site index $x$ and a gauge index $a$ (before, we had bundled them together into a single multi-index $i$). On the quantum Hilbert space gauge transformations are implemented at each site by a unitary operator $\Gamma(g)$ which satisfies
	\begin{align}
		\Gamma(g) \psi_{a} \Gamma(g)^\dagger &=\rho(g^{-1})_{ab}\psi_b \ ,\\
		\Gamma(g) \psi_{a}^\dagger \Gamma(g)^\dagger &=\psi_b^\dagger \rho(g)_{ba} \ .
	\end{align}
	Since $\rho$ is a unitary group representation, the two operators are indeed adjoint to each other. This choice of transformation rule ensures that the states $\psi_a^\dagger \ket{\Omega}$ indeed transform according to the representation $\rho$ (as can be checked explicitly), consistently with their identification with $e_a \in V_F$.
	
	The action of $\Gamma$ on an arbitrary state in $\mathcal{H}_F$ can be computed as (suppressing $g$)
	\begin{align}
		\Gamma \pqty{\psi_{a_1}^\dagger \psi_{a_2}^\dagger \cdots \psi_{a_k}^\dagger \ket{\Omega}} & = \pqty{\Gamma \psi_{a_1}^\dagger \Gamma^\dagger} \pqty{\Gamma \psi_{a_2}^\dagger \Gamma^\dagger} \cdots (\Gamma\psi_{a_k}^\dagger \Gamma^\dagger)\Gamma \ket{\Omega} = \label{eq:Gamma action on basis state}\\
		&= \rho_{b_1 a_1} \rho_{b_2 a_2 } \cdots \rho_{b_k a_k} \psi_{b_1}^\dagger \psi_{b_2}^\dagger \cdots \psi_{b_k}^\dagger \Gamma \ket{\Omega} \nonumber \ .
	\end{align}
	Note that it is in fact consistent that the vacuum state transforms as an arbitrary one-dimensional representation $\sigma$ of the gauge group $G$,
	\begin{equation}
		\Gamma(g) \ket{\Omega} = \sigma(g) \ket{\Omega} \ .
	\end{equation}
	As we will see, it will be necessary to consider this possibility in some cases. For now, we assume that the vacuum is invariant under $\Gamma$. Then note that $\Gamma$ maps the rank $k$ subspace of $\mathcal{H}_F$ to itself; on this subspace it acts as the totally antisymmetric representation. In the exterior algebra language, this is known as the $k$th \textit{exterior power} $\Lambda^k \rho$ of $\rho$, which acts (by definition) on the subspaces $\Lambda^k V_F$ of rank $k$ pure wedges as
	\begin{equation}
		\label{eq:kth exterior power representation action}
		(\Lambda^k\rho)(g)(v_{a_1} \wedge v_{a_2} \wedge \cdots \wedge v_{a_k}) = (\rho(g) v_{a_1}) \wedge (\rho(g)v_{a_2}) \wedge \cdots \wedge (\rho(g) v_{a_k}) \ , 
	\end{equation}
	which under the identification $\mathcal{H}_F \cong \Lambda(V_F)$ is the same as the action of $\Gamma$ eq.\eqref{eq:Gamma action on basis state}. Therefore $\Gamma$, as a representation of the gauge group $G$ on $\mathcal{H}_F$, is isomorphic to the \textit{exterior algebra representation} $\Lambda\rho$ on $\Lambda(V_F)$,
	\begin{equation}
		\Gamma \cong \Lambda \rho \equiv \bigoplus_{k=0}^N \Lambda^k \rho \ .  
	\end{equation}
	
	As is clear from the discussion in Section \ref{sec:master counting}, the key element for the computation of the number of gauge-invariant states is the character (i.e. the trace) of the representation of the gauge group on the matter Hilbert space. Thus to make progress we need to compute the character $\chi_{\Lambda\rho}$ of $\Lambda \rho$. This can be done as follows. First of all, we have
	\begin{equation}
		\label{eq:exterior algebra character decomposition}
		\chi_{\Lambda\rho} = \sum_{k=0}^N \chi_{\Lambda^k\rho} \ .
	\end{equation}
	Since it will be useful later, and is useful for bookkeeping, consider the following polynomial of a real variable $t$ (for fixed $g$): 
	\begin{equation}
		\label{eq:P(t) definition}
		P(t) \equiv \sum_{k=0}^N \chi_{\Lambda^k\rho}(g) t^k \ .
	\end{equation}
	Then $\chi_{\Lambda \rho}(g) \equiv P(1)$. To make progress, we compute the character $\chi_{\Lambda^k\rho}$. Since $\rho(g)$ is unitary, it is diagonalizable, so it has eigenvalues $\lambda_i$ for $i=1,\ldots , N$ (i.e. counted with multiplicity) associated to eigenvectors $v_i \in V_F$, i.e. $\rho(g) v_i = \lambda_i v_i$ (no sum). The eigenvectors form an orthonormal basis of $V_F$. Thus, as we saw not long ago (eq.\eqref{eq:identification basis HF Lambda}), an orthonormal basis of $\Lambda^k (V_F)$ is given by the pure wedges
	\begin{equation}
		v_{i_1} \wedge v_{i_2 } \wedge \cdots \wedge v_{i_k} \ ,
	\end{equation}
	for $1 \leq i_1 < i_2 < \cdots < i_k \leq N$. Then the action of $(\Lambda^k\rho)(g)$ on this basis is given by (see eq.\eqref{eq:kth exterior power representation action})
	\begin{align}
		(\Lambda^k\rho)(g)(v_{i_1} \wedge v_{i_2} \wedge \cdots \wedge v_{i_k}) &\equiv (\rho(g) v_{i_1}) \wedge (\rho(g)v_{i_2}) \wedge \cdots \wedge (\rho(g) v_{i_k})=\\
		&=\lambda_{i_1} \lambda_{i_2} \cdots \lambda_{i_k} (v_{i_1} \wedge v_{i_2} \wedge \cdots \wedge v_{i_k}) \ .
	\end{align}
	Therefore in this basis $(\Lambda^k\rho)(g)$ is diagonal and its eigenvalues can be read off from the previous equation. Therefore its trace is given by
	\begin{equation}
		\chi_{\Lambda^k\rho}(g) \equiv \tr (\Lambda^k\rho)(g) = \sum_{1 \leq i_1 < i_2 < \cdots < i_k \leq N} \lambda_{i_1} \lambda_{i_2} \cdots \lambda_{i_k} \ .
	\end{equation}
	This expression is also known as the $k$th elementary symmetric polynomial on $N$ variables $\lambda_1, \lambda_2, \ldots, \lambda_N$. Substituting into the polynomial, we find
	\begin{equation}
		\label{eq:P(t) expansion}
		P(t) = \sum_{k=0}^N t^k \sum_{1 \leq i_1 < i_2 < \cdots < i_k \leq N} \lambda_{i_1} \lambda_{i_2} \cdots \lambda_{i_k} \ .
	\end{equation}
	The coefficient of $t^k$ in $P(t)$ is nothing but the sum of all possible products of variables $\lambda_i$ with $k$ different indices. But this then means that $P(t)$ can be factored as
	\begin{equation}
		\label{eq:P(t) factorization}
		P(t) = \prod_{i=1}^N (1+\lambda_i t) \ .
	\end{equation}
	Indeed, expanding eq.\eqref{eq:P(t) factorization} one obtains precisely eq.\eqref{eq:P(t) expansion}. This last expression is nothing but a determinant, i.e.
	\begin{equation}
		\label{eq:P(t) determinant}
		P(t) \equiv \det(\mathds{1}+\rho(g)t) \ .
	\end{equation}
	Therefore this means that
	\begin{equation}
		\label{eq:character of exterior algebra representation}
		\chi_{\Lambda \rho}(g) = \det(\mathds{1}+\rho(g)) \ .
	\end{equation}
	Using this expression, one can show explicitly that the group representation factors as a tensor product, i.e.
	\begin{equation}
		\label{eq:exterior algebra representation factorization}
		\Lambda(\rho_1\oplus\rho_2) \cong \Lambda(\rho_1) \otimes \Lambda(\rho_2) \ ,
	\end{equation}
	consistently with the decomposition eq.\eqref{eq:exterior algebra direct sum property}. This is because if $n_i = \dim \rho_i$, then
	\begin{align}
		\chi_{\Lambda(\rho_1 \oplus\rho_2)} &= \det(\mathds{1}_{n_1+n_2} +\rho_1 \oplus \rho_2) = \det\begin{pmatrix}
			\mathds{1}_{n_1} + \rho_1 & 0\\
			0 & \mathds{1}_{n_2} + \rho_2 
		\end{pmatrix} =\\
		&=\det{(\mathds{1}_{n_1}+\rho_1)} \det{(\mathds{1}_m+\rho_2)} = \chi_{\Lambda(\rho_1)} \chi_{\Lambda(\rho_2)} \ .
	\end{align}
	As mentioned at the beginning of this section, the fermion field may carry additional spinor and/or flavour indices. Each additional flavour appears in the Hilbert space as a direct summand of the form $(V_x)_{\mathrm{spin}} \otimes (V_x)_{\mathrm{gauge}}$, so they can be dealt with using the direct sum formula above. On the other hand, the gauge group acts on the spinor degrees of freedom as the identity operator, so it acts on $V_x$ as $\mathds{1}_{N_s} \otimes \rho$ where $N_s$ is the number of spinor indices. Then if $n=\dim{\rho}$, one sees that
	\begin{align}
		\chi_{\Lambda(\mathds{1}_{N_s}\otimes\rho)} &= \det{(\mathds{1}_{N_s n}+\mathds{1}_{N_s} \otimes \rho_1)} = \\
		&=\det \begin{pmatrix}
			\mathds{1}_n + \rho &  &  &  & \\
			& \mathds{1}_n + \rho &  &  & \\
			& & \ddots & \\
			&  & & \mathds{1}_n + \rho &
		\end{pmatrix} =\\
		&=\pqty{\det{(\mathds{1}_n + \rho)}}^{N_s} = (\chi_{\Lambda \rho} )^{N_s}
	\end{align}
	Thus we simply exponentiate the previous result by the number of spinor indices.
	
	Using the general result eq.\eqref{eq:master counting formula} from Section \ref{sec:master counting}, we therefore have the following result. Suppose the fermions have $N_s$ spinor indices and assume that the fermions transform in the same representation $\rho$ at all sites $x$. Assume further that the Fock vacuum transforms under the one-dimensional representation $\sigma$. If the gauge transformations are the same at all sites (i.e. no twisted boundary conditions as in Section \ref{sec:twisted}), then on a lattice with $V$ sites and $E$ links, the number of gauge-invariant states is
	\begin{equation}
		\label{eq:fermion gauge-invariant states number}
		\dim \mathcal{H}_\mathrm{phys} = \sum_{C} \pqty{\frac{\abs{G}}{\abs{C}}}^{E-V} \sigma(C)\det{(\mathds{1} + \rho(C))}^{N_s  V} \ ,
	\end{equation}
	where the sum is over all conjugacy classes $C$ of $G$. Note that the fermion character vanishes on all conjugacy classes $C$ where $\rho(g)$ (for $g \in C)$ has an eigenvalue equal to $-1$. Note further that the determinant term is always non-negative, and since at $C=1$ it takes the value $2^{\dim \rho}$, the number of gauge-invariant states is never zero. If one has multiple fermion flavours, then eq.\eqref{eq:fermion gauge-invariant states number} gets an additional determinant factor per flavour, each with a possibly different group representation. For completeness, the size of the total Hilbert space is
	\begin{equation}
		\dim \mathcal{H}_\mathrm{tot} = \abs{G}^E 2^{N_s N_f V \dim \rho} \ .
	\end{equation}
	Here $N_f$ is the number of flavours. For example, for the dihedral group $D_4$ of order $8$, if we take the fermions to lie in the two-dimensional (faithful) irreducible representation, one finds the formula
	\begin{equation}
		\dim \mathcal{H}_\mathrm{phys} = 8^{E-V} 4^{N_s N_f V} + 8^{E-V} 2^{N_s N_fV} + 4^{E-V} 2^{N_s N_fV} \ .
	\end{equation}
	In the following subsection we discuss how to adapt the general formula for fermions to the cases of naive, Wilson and staggered fermions. We then conclude by discussing fermion parity superselection and how it can be incorporated.
	
	\subsection{Naive and Wilson fermions}
	
	Wilson fermions represent one of the possible solutions to the fermion doubling problem \cite{Wilson1974}. The free Hamiltonian is given by
	\begin{equation}
		H_M = a^d\sum_{x} \Psi^\dagger(x) \gamma_0 \bqty{i \gamma_i \partial_i + m + ar\Delta} \Psi(x) \ ,
	\end{equation}
	where (in $3+1$ dimensions) $\Psi$ is a four-component Dirac spinor which also carries a gauge index and possibly a flavour index; its components satisfy canonical commutation relations. More generally, a Dirac spinor has $N_s = 2^{\lfloor (d+1)/2 \rfloor}$ in $d$ space dimensions.
	
	The discrete (lattice) derivative $\partial_i$ and the lattice Laplacian $\Delta$ are normalized by factors of the lattice spacing $a$. On the other hand, $r$ is a parameter; Wilson fermions correspond to $0 < r \leq 1$ while naive fermions correspond to $r=0$. In particular, the two formulations share the same Hilbert space: it is the Fock space generated by the components of $\Psi$ as described in the previous section. The same conclusion can be reached by the transfer-matrix method \cite{Luscher:1976ms, CreutzTransferMatrix}.
	
	Writing out the various terms more explicitly, one has
	\begin{equation}
		H_M = a^{d-1} \sum_x \Psi^\dagger(x) \gamma_0 \bqty{ \tfrac12\pqty{i\gamma_{i} -r } \Psi(x + a\hat{i}) - \tfrac12\pqty{i \gamma_{i}+r} \Psi(x-a\hat{i})+ \pqty{a m+r} \Psi(x) } \ .
	\end{equation}
	This Hamiltonian can then be coupled to the gauge field by the standard prescription. One introduces a gauge field $U_{x,i}$ in the same representation as the fermions:
	\begin{equation}
		H_M = a^{d-1} \sum_x \Psi^\dagger(x) \gamma_0 \bqty{ \tfrac12\pqty{i\gamma_{i} -r } U_{x,i}^\dagger \Psi(x + a\hat{i}) - \tfrac12\pqty{i \gamma_{i}+r} U_{x,i}\Psi(x-a\hat{i})+ \pqty{a m+r} \Psi(x) } \ .
	\end{equation}
	The gauge field is made dynamical by the addition of its Hamiltonian eq.\eqref{eq:gauge hamiltonian}. 
	
	Therefore, for the purpose of computing the number of gauge invariant states, one uses the formula eq.\eqref{eq:fermion gauge-invariant states number} with $N_s = 2^{\lfloor (d+1)/2 \rfloor}$ in $d$ space dimensions. The fermionic vacuum is invariant, thus $\sigma = 1$.
	
	\subsection{Staggered fermions}
	
	Free staggered fermions, on the other hand, are given by the Hamiltonian \cite{SusskindFermions, Catterall:2025vrx}
	\begin{equation}
		H_M = \sum_x \chi^\dagger (x) [i \eta_i(x) \partial_i + (-1)^x m ] \chi(x) \ ,
	\end{equation}
	where $\eta_i(x) = (-1)^{\sum_{i<j}x_i}$ on a hypercubic lattice. Importantly, $\chi$ is a single-component fermion which satisfies canonical anticommutation relations. Similarly to the case of Wilson fermions, the fermion field carries a gauge index and possibly a flavour index; gauge fields are introduced similarly.
	
	Note however an important difference. Due to the staggered phase factors, the ground state of the mass term is itself \say{staggered}: it is the empty Fock vacuum for even sites (i.e. $\chi^\dagger \chi$ is zero), while it is the maximally occupied state for odd sites (i.e. $\chi^\dagger \chi$ is as large as possible). It has been argued \cite{ZoharBurrello} that this fact should be reflected in the Gauss law. In other words, the vacuum on even sites should not transform under gauge transformations, while the vacuum on odd sites should transform like a full site, i.e. with a determinant factor. Summing up, the vacuum should transform under gauge transformations via a one-dimensional representation $\sigma_x$ given by 
	\begin{equation}
		\sigma_x(g) = \begin{cases} 1 & x\,\,\mathrm{even}\\ \det{\rho(g^{-1})} & x\,\,\mathrm{odd} \end{cases} \ .
	\end{equation}
	Note that the presence of this factor does not affect the transformation law of the fermionic operators.
	
	Overall, for the computation of the number of gauge-invariant states for staggered fermions, one uses again eq.\eqref{eq:fermion gauge-invariant states number} with $N_s=1$ and (assuming $V$ even)
	\begin{equation}
		\sigma(C) = \det{\rho(C^{-1})}^{V/2} \ .
	\end{equation}
	
	\subsection{Fermion parity superselection}\label{sec:superselection}
	
	As is clear from eq.\eqref{eq:Gamma action on basis state}, gauge transformations preserve the number of creation operators on each site. At each site $x$, fermion number $F(x)$ is given by
	\begin{equation}
		F(x) = \sum_{a} \psi_a^\dagger(x)\psi_a(x) \ ,
	\end{equation}
	and it commutes with gauge transformations, $[F(x), \Gamma]=0$. A particularly important operator is total fermion parity $(-1)^F$ where $F = \sum_x F(x)$. 
	
	Basic principles imply the existence of a superselection rule which forbids the superposition of fermionic and bosonic states \cite{Wick:1952nb,Wick:1970bd,johansson2016}. The superselection rule can also be stated by saying that physical observables cannot change fermion number. For these reasons, one may want to split $\Hphys$ further into eigenspaces of $(-1)^F$. Here we show how to compute the number of gauge-invariant states in each eigenspace. To this end, note that the projectors onto the two $\pm 1$ eigenspaces are given by
	\begin{equation}
		P_{\pm} =\frac{1\pm (-1)^F}{2} \ .
	\end{equation}
	Then, following the reasoning in Section \ref{sec:master counting}, the number of gauge-invariant states in each eigenspace is given by $\tr(P P_{\pm})$ where $P$ is the projector onto $\Hphys$ given in eq.\eqref{eq:projector onto Hphys}. Note that the non-trivial aspect of this calculation is the computation of $\tr(\Gamma(-1)^F)$. But note that fermion parity acts on $\Lambda^kV$ as $(-1)^k$, so that one has, similarly to eq.\eqref{eq:exterior algebra character decomposition}
	\begin{equation}
		\tr(\Gamma(-1)^F) = \sum_{k=0}^N (-1)^k \chi_{\Lambda^k\rho} = P(-1) \ ,
	\end{equation}
	where $P$ is the polynomial defined in eq.\eqref{eq:P(t) definition}. From the general result eq.\eqref{eq:P(t) determinant}, we therefore find that
	\begin{equation}
		\tr(\Gamma(g)(-1)^F) = \det(\mathds{1}-\rho(g)) \ .
	\end{equation}
	From this, one can then assemble the full answer in each case of interest.
	
	\section{Conclusion}
	
	In this work, we have shown how to compute the number of gauge-invariant states for lattice gauge theories with finite gauge group and arbitrary matter fields, as well as some types of twisted boundary conditions. This completes and extends the same calculation which had been performed in the case of pure gauge theories \cite{MPE, Mariani:2024osg}. These formulas have applications to resource estimation for quantum simulation, as well as more generally whenever one works with gauge-invariant methods.
	
	We have discussed the calculation in detail in several cases of interest. This does not exhaust all useful possibilities, but we hope to have provided the reader with a sufficiently powerful toolkit to address this question whatever their relevant situation. We briefly summarize the various cases for which we've obtained formulas for the number of gauge-invariant states. In particular, we have given a master formula eq.\eqref{eq:master counting formula} which applies as long as one does not have twisted boundary conditions. This formula was instantiated in the case of scalar matter in eq.\eqref{eq:counting formula scalars} and for fermions in eq.\eqref{eq:fermion gauge-invariant states number}. For certain types of twisted boundary conditions, we have instead obtained eq.\eqref{eq:master formula twisted} and eq.\eqref{eq:master formula twisted non homomorphism}, which may also be combined with the results for scalars and fermions. In case matter fields transform under different representations in different sites, one can instead use eq.\eqref{eq:master formula different sites}; for the dimension of non-trivial Gauss law sectors, instead one can use eq.\eqref{eq:counting formula charged sectors}. Finally, we have given an example of how one can incorporate further constraints into the count in the case of fermion parity in Section \ref{sec:superselection}.
	
	As part of our discussion of twisted boundaries, we have pointed out that some symmetries in the gauge sector are implemented by outer automorphisms of the gauge group, among which is charge conjugation. In particular, we point out that while $\SU(N)$ admits an unambiguous definition of charge conjugation, this is not the case for all of its finite subgroups. We have checked that this does not pose a practical issue in most cases, but it is an important fact to keep in mind.
	
	It would be interesting to extend these results to other methods for truncating the infinite-dimensional Hilbert space of gauge theories. Unfortunately, at least for two of the methods familiar to the author (truncation in the representation basis and quantum link models) it does not appear to be possible to obtain a simple formula such as the one we obtained for finite groups, even restricting to pure gauge theories. In the former case, an exact formula can be obtained in principle by the methods of \cite{MPE}, but it does not appear to simplify due to the lack of group-theoretical identities which are lost when considering only a subset of representations. In the latter case (quantum link models), even just for the spin $1/2$ $\U(1)$ quantum link model the problem in general is equivalent to an open problem in graph theory \cite{Beck2006}, and is in fact closely related to an intractable (i.e. \texttt{\#P-hard}) computational problem \cite{BaldoniSilva2004}. While it may still be possible to obtain simple formulas in special cases, the problem seems significantly harder. We will discuss these issues in more detail in a forthcoming publication.
	
	Another interesting case would be to consider chiral fermions, especially with respect to anomalies. Unfortunately, there is to this day no satisfactory Hamiltonian formulation of chiral gauge theories (see e.g. \cite{Singh:2025sye}); moreover, in this case one expects that gauge transformations do not act ultralocally, which would make our methods inapplicable.

	\section*{Acknowledgments}
	
	The author thanks Uwe-Jens Wiese for reading the manuscript. The author is supported by the Simons Foundation grant 994300 (Simons Collaboration on Confinement and QCD Strings). Support from the SFT Scientific Initiative of the Italian Nuclear Physics Institute (INFN) is also acknowledged.
	
	\printbibliography
	
	\appendix

	\section{Some results on group actions}\label{sec:group actions}
	
	As we have seen in Section \ref{sec:scalar}, group actions describe the classical symmetry of scalar field theories. However, as we have seen in Section \ref{sec:scalar examples} if the actions satisfy certain common properties, then they are equivalent to certain special actions:
	\begin{enumerate}
		\item The action of $G$ on a coset space $G/H$ by left multiplication is the general case of a transitive action (an action of $G$ on $S$ is transitive if, given any two $s, s' \in S$, there is $g \in G$ such that $s'=g\cdot s)$. 
		\item The action of $G$ on a set $G \times \Sigma$ by left multiplication on the first factor is the general case of a free action (an action is \textit{free} if it has no fixed points, i.e. $g\cdot s=s$ for \textit{any} $s$ implies $g=1$). 
		\item The action of $G$ on itself by left multiplication is the general case of an action that is both transitive and free (such an action is also called \textit{regular}, \textit{simply transitive} or \textit{sharply transitive}). 
	\end{enumerate}
	
	In the rest of this section we prove these results, all of which are well-known in the mathematical literature. The proofs give explicit constructions of the isomorphism. 
	
	The first result is about transitive actions:
	
	\begin{theorem}
		A transitive action of a finite group $G$ on a set $S$ is isomorphic to the action of $G$ by left multiplication on the coset space $G/H$ for some subgroup $H \leq G$.
	\end{theorem}
	
	\begin{proof}
		Choose an arbitrary element $s_* \in S$ and let $H = \mathrm{Stab}(s_*)$, the stabilizer of $s_*$. Since by assumption $G$ acts transitively on $S$, given any $s \in S$ there is a (non-unique) element $g_s$ such that $s = g_s \cdot s_*$. Now define a map $f: S \to G/H$ by sending $f(s)=g_s H$. We first need to show that this is well defined (since $g_s$ is not unique, $f(s)$ may a priori depend on which $g_s$ is chosen). Suppose that $s = g_s \cdot s = g_s' \cdot s$. Then $g_s^{-1} g_s' \in \mathrm{Stab}(s_*) = H$, which then implies that $g_s H = g_s' H$, i.e. $f(s)$ does not depend on which $g_s$ is picked. 
		
		Now we show that in fact $f$ is a bijection. Injectivity follows by reversing the order of the statements of the last proof. For surjectivity, given a coset $x \in G/H$ pick a representative $g$ and compute $g\cdot s_*$ (since $H=\mathrm{Stab}(s_*)$, the result does not depend on the choice of representative). Then $f(g\cdot s_*)=x$.
		
		It remains to show that the action of $G$ on $S$ is isomorphic to the action of $G$ on $G/H$ by left multiplication. We have
		\begin{equation}
			f(g\cdot s) = f(g g_s \cdot s_*) = g g_s H = g f(s) \ ,
		\end{equation}
		which proves the result. Note that the proof is constructive.
	\end{proof}
	
	This result is also valid for compact Lie groups as long as the action is smooth. The quotient space $G/H$ then is a manifold.
	
	Now we consider the second case, that of a free action. 
	
	\begin{theorem}
		If a finite group $G$ acts on a set $S$ freely, then the action is isomorphic to the action of $G$ on $G \times \Sigma$ for some other set $\Sigma$ by left multiplication on the first factor.
	\end{theorem}
	
	\begin{proof}
		The key observation is that since the action is free, then all stabilizers are trivial. In fact by definition of a free action, if $g \cdot s =s$ for any $s$ then $g=1$.
		
		Let $\Sigma = S/G$, i.e. the set of orbits of $S$ under the free action of $G$. For each orbit $x \in \Sigma$, pick a representative $s_x \in S$. If $s \in x$ is any other element of the same orbit, then $s = g_s s_x$ for some $g_s \in G$ because they lie in the same orbit. Moreover, since the action is free, $g_s$ is unique. Now having fixed a set of representatives for each orbit, define a map $f: S \to G \times \Sigma$ by setting 
		\begin{equation}
			f(s) = (g_s, [s]) \ ,
		\end{equation}
		where $[s] \in \Sigma$ is the orbit of $s$. This map is a bijection, because we can construct an inverse map $f^{-1}: G \times \Sigma \to S$ explicitly by setting $f^{-1}(g, x)=g \cdot s_x$.  
		
		It remains to show that the action of $G$ on $G \times \Sigma$ is given by left multiplication on the first factor. We have:
		\begin{equation}
			f(g \cdot s) = f( gg_s \cdot s_x) = (gg_s, x) = g \cdot (g_s, x) \ .
		\end{equation}
		Again the proof is constructive.
	\end{proof}
	
	In this case, the key observation is that all stabilizers are trivial and therefore all orbits are isomorphic to $G$. This last fact is also true for smooth actions of a compact Lie group.
	
	The last result follows from the first:
	
	\begin{theorem}
		A transitive and free action of a finite group $G$ on a set $S$ is isomorphic to the action of $G$ on itself by left multiplication.
	\end{theorem}
	
	\begin{proof}
		By the first theorem of this appendix, we know that the action (being transitive) is isomorphic to the action of $G$ on $G/H$ by left multiplication. By the proof, we know that $H$ is the stabilizer of any element. But since the action is free, all stabilizers are trivial and therefore $H$ is the trivial group, so $G/H \cong G$ and the result follows.
	\end{proof}
	
\end{document}